\documentclass[%
reprint, 
prx,
amsmath,amssymb,
aps,
floatfix,
]{revtex4-2}

\usepackage{graphicx}
\usepackage{physics}
\usepackage{bm} 

\usepackage{booktabs}
\usepackage{caption}
\usepackage{subcaption}
\usepackage{makecell} 

\usepackage[colorlinks, allcolors=blue]{hyperref}

\def\urbana{
The Anthony J. Leggett Institute for Condensed Matter Theory and IQUIST and NCSA Center for Artificial Intelligence Innovation and Department of Physics, University of Illinois at Urbana-Champaign, IL 61801, USA}

\begin{document}
\title{Efficient optimization of neural network backflow for ab-initio quantum chemistry}

\author{An-Jun Liu} 
\affiliation{\urbana}
\author{Bryan K. Clark}
\affiliation{\urbana}

\begin{abstract}
The ground state of second-quantized quantum chemistry Hamiltonians is key to determining molecular properties. Neural quantum states (NQS) offer flexible and expressive wavefunction ansatze for this task but face two main challenges: highly peaked ground-state wavefunctions hinder efficient sampling, and local energy evaluations scale quartically with system size, incurring significant computational costs. In this work, we overcome these challenges by introducing a suite of algorithmic enhancements, which includes efficient periodic compact subspace construction, truncated local energy evaluations, improved stochastic sampling, and physics-informed modifications.

Applying these techniques to the neural network backflow (NNBF) ansatz, we demonstrate significant gains in both accuracy and scalability. Our enhanced method surpasses traditional quantum chemistry methods like CCSD and CCSD(T), outperforms other NQS approaches, and achieves competitive energies with state-of-the-art ab initio techniques such as HCI, ASCI, FCIQMC, and DMRG. A series of ablation and comparative studies quantifies the contribution of each enhancement to the observed improvements in accuracy and efficiency. Furthermore, we investigate the representational capacity of the ansatz, finding that its performance correlates with the inverse participation ratio (IPR), with more delocalized states being more challenging to approximate.
\end{abstract}
\maketitle

\section{Introduction}
Accurately solving the many-electron Schrödinger equation is central to quantum chemistry (QC) and condensed matter physics, as knowledge of a system’s ground state enables the prediction of a wide range of physical and chemical properties from first principles. However, this problem is fundamentally NP-hard \cite{Whitfield2013,Gorman2022}, which has driven the development of numerous approximation methods. Instead of directly computing the full eigenspectrum of the Hamiltonian, variational approaches focus on minimizing the expected energy of a parameterized wavefunction ansatz. Over the decades, many such ansatze have been proposed, including Configuration Interaction (CI) methods \cite{DavidSherrill1999}, Coupled Cluster (CC) techniques \cite{Coester1960}, Slater–Jastrow (SJ) forms \cite{Foulkes2001,Clark2011}, Matrix Product States (MPS) \cite{White1992,White1999}, and Selected Configuration Interaction (SCI) methods \cite{Holmes2016,Sharma2017,Tubman2016,Tubman2018}.

In recent years, machine learning has emerged as a powerful tool for constructing concise, flexible, and expressive wavefunction ansatze. Neural quantum states (NQS) \cite{Carleo2017} leverage the ability of neural networks to represent complex, high-dimensional probability distributions.  While NQS were originally focused on spin models \cite{Choo2019,Choo2018,Nagy2019,Sharir2020}, more recently, there has been a growing effort to extend these methods to fermionic systems \cite{Nomura2017,Francesco2019,Stokes2020,Lin2023,Zhuo2022,Di2019,zejun2023,Pfau2020,Hermann2020,Choo2020,Barrett2022,Zhao2023,Li2023,Shang2023,Wu2023,Malyshev2023,Liu2024,Li2024,Malyshev2024,Knitter2024}.

Starting with the work of Choo et. al~\cite{Choo2020}, NQS have also been applied to molecular Hamiltonians in a second quantized formalism \cite{Barrett2022,Zhao2023,Li2023,Shang2023,Wu2023,Malyshev2023,Liu2024,Li2024,Malyshev2024,Knitter2024}.  The neural network backflow (NNBF) ansatz \cite{Liu2024} is one of the most accurate approaches to second quantized QC Hamiltonians and consistently achieves state of the art results.  
Despite this progress, two significant challenges remain both for NNBF and other NQS ansatz in this space. 
First, molecular ground-state wavefunctions often exhibit a highly peaked structure, dominated by a few high-amplitude configurations. This poses a major challenge for sampling the Born distribution using standard Markov chain Monte Carlo (MCMC) methods—the default in many VMC implementations. Second, while the number of terms in local energy evaluations grows polynomially—specifically at a quartic rate—with system size, these computations become progressively more demanding for larger systems, posing a significant challenge for NQS applications.

Various strategies have been developed to mitigate these issues. Autoregressive neural networks offer exact and efficient sampling capabilities, bypassing MCMC’s limitations \cite{Barrett2022,Zhao2023}. Deterministic selection approaches \cite{Li2023,Liu2024} reduce reliance on stochastic sampling, and alternative techniques have been proposed to streamline local energy computations \cite{Wu2023}. Although these methods have alleviated certain bottlenecks, most state-of-the-art results remain confined to relatively small systems where exact diagonalization remains feasible. Thus, there is still a need for compelling evidence that NQS approaches can scale to more challenging molecular systems while delivering competitive energy accuracies at larger system sizes.

In this work, building upon our previous study \cite{Liu2024}, we introduce a suite of algorithmic advancements that significantly improve the accuracy and scalability of Neural Quantum State (NQS) methods for quantum chemistry. These enhancements, which form a general and robust optimization framework, consist of four main components: an efficient method for constructing and periodically updating a compact yet important subspace; a truncated local energy evaluation strategy reusing pre-computed information; an improved stochastic sampling method to provide more unbiased energy estimations; and the incorporation of prior physical knowledge into the ansatz and training pipeline.

We apply this framework to the NNBF ansatz and benchmark its performance on various challenging systems, including the paradigmatic strongly correlated N$_2$ molecule. Our results show that the enhanced NNBF method not only continues to outperform traditional methods like CCSD and CCSD(T) and all other existing NQS approaches but also achieves competitive performance against state-of-the-art ab-initio methods such as HCI \cite{Holmes2016,Sharma2017}, ASCI \cite{Tubman2016,Tubman2018}, FCIQMC \cite{Cleland2012}, and DMRG \cite{Chan2004}. To quantify the contribution of each proposed improvement, we conduct an ablation study that cumulatively adds these techniques. The results show that our enhanced approach achieves orders-of-magnitude improvements in energy accuracy while substantially reducing the wall clock time per optimization step. To provide deeper insight, this is followed by two focused studies: one dissecting the components of our stochastic sampling method, and another comparing our local energy evaluation strategy against common alternatives. Finally, we investigate how the expressiveness of the NNBF ansatz depends on the inverse participation ratio (IPR) of the target quantum state, providing additional insights into the factors influencing the representational capacity of NQS.
\section{Methods}

In this section, we first present an overview of the background and previous works, followed by a detailed description of the algorithmic improvements proposed in this study.

\begin{figure*}[t]\label{fig:algo_sum}
    \centering
    \includegraphics[width=\linewidth]{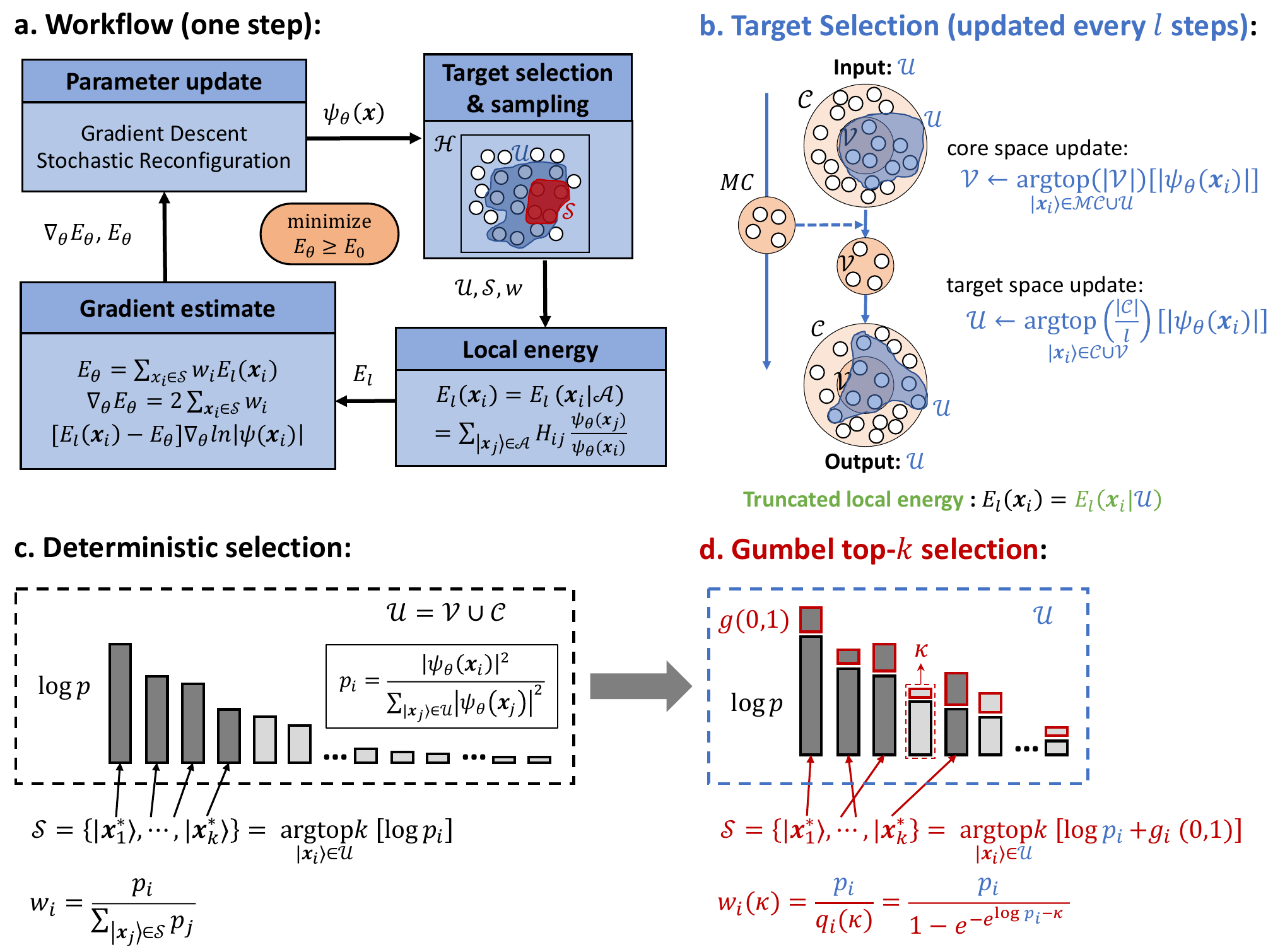}
    \caption{
    (a) The Variational Monte Carlo (VMC) workflow. Each optimization iteration consists of four subroutines:
    (upper right) Identification (every $l$ iterations) of configuration in the target space $\mathcal{U}$ within the full Hilbert space $\mathcal{H}$; a sample set $\mathcal{S}$ will be drawn from $\mathcal{U}$ at each optimization step and corresponding importance weights $w_i$ are assigned; 
    (lower right) The local energies $E_l(\mathbf{x}_i|\mathcal{A})$ are computed for each sample $\mathbf{x}_i \in \mathcal{S}$ given a subspace $\mathcal{A}$, where $\mathcal{A}$ restricts the set of configurations considered when iterating over all connected configurations expanded from $\ket{\mathbf{x}_i}$. Examples of $\mathcal{A}$ are the full Hilbert space $\mathcal{H}$ (when computed exactly) or the target space $\mathcal{U}$;
    (lower left) The total energy and its gradient are estimated via an importance-sampled average using the samples $\mathcal{S}$, weights $w_i$, and local energies $E_l$.
    (upper left) An optimizer, such as gradient descent or Stochastic Reconfiguration, uses these estimates to update the model parameters.
    (b) The Intermittent Target Selection (ITS) update, performed every $l$ steps. First, the core space $\mathcal{V}$ is updated by selecting the $\abs{\mathcal{V}}$ highest-amplitude unique configurations from the union of the prior target space $\mathcal{U}$ and a set of $\abs{\mathcal{V}}$ MCMC walkers maintained concurrently to provide access to the current wave-function. Subsequently, a new target subspace $\mathcal{U}$ is constructed by selecting the $\abs{\mathcal{C}}/l$ configurations with largest amplitude moduli from the updated core space and its newly generated connected space. This subspace $\mathcal{U}$ then remains fixed until hitting the next ITS step, with all amplitude calculations being restricted to this compact yet important subspace $\mathcal{U}$.
    (c) Workflow of the Fixed-Size Selected Configuration (FSSC) method \cite{Liu2024} for comparison. The target space is the union of the core and connected spaces, i.e., $\mathcal{U} = \mathcal{V} \cup \mathcal{C}$. The sample is deterministically constructed by selecting the $k$ unique configurations with the largest amplitude magnitudes, and the importance weights are probabilities renormalized with respect to this sample. The local energy is computed exactly.
    (d) Workflow for the stochastic sampling via the Gumbel top-$k$ trick used in the current work  . The log-probabilities $\log{p_i}$ of configurations in the target space $\mathcal{U}$ are perturbed with Gumbel noise $g_i \overset{i.i.d.}{\sim} \text{Gumbel}(0,1) \quad \text{with pdf} \quad f(x)=e^{-(x+e^{-x})}$. The sample $\mathcal{S}$ is then formed by selecting the $k$ configurations with the largest perturbed log-probabilities. The corresponding importance weights are calculated to provide unbiased estimates over the subspace $\mathcal{U}$, using a threshold $\kappa$ defined as the value of the $(k+1)$-th largest perturbed log-probability.
    }
\end{figure*}

\subsection{Overview}\label{sec:overview}

For a many-body system containing $N_e$ electrons and $N_o$ single-particle orbitals (SPOs) $\mathcal{B}=\{ \phi_i \}_{i=1}^{N_o}$, the many-electron wavefunction can be expressed in second quantization as $\ket{\psi} = \sum_{i}\psi(\mathbf{x}_i)\ket{\mathbf{x}_i}$ where $\ket{\mathbf{x}_i}=\ket{x_i^{1\uparrow},\dots,x_i^{N_o\uparrow},x_i^{1\downarrow},\dots,x_i^{N_o\downarrow}}$ is the i-th computational basis vector, with $x_i^j \in \{0,1\}$ denoting the occupation of the j-th spin-orbital. NQS offer a promising framework for efficiently representing many-electron wavefunctions via machine learning architectures: $\ket{\psi_\theta} = \sum_{i}\psi_\theta(\mathbf{x}_i)\ket{\mathbf{x}_i}$ where $\theta$ are model parameters. Among various NQS architectures, NNBF has demonstrated strong performance in fermionic systems \cite{Di2019,zejun2023,Zhuo2022,Liu2024}. The NNBF wavefunction is defined as $\psi_\theta(\mathbf{x}_i)=\sum_{m=1}^{D}\det[\Phi^m_{j=\{l | x_i^l=1\},k}(\mathbf{x}_i;\theta)]$ where $\Phi^m_{jk}$ are ``configuration-dependent'' spin-orbitals output by NNBF’s internal network.

To determine the ground state of quantum systems, rather than solving the electronic Schrödinger equation directly, VMC reformulates it as an optimization problem by minimizing the variational energy
\begin{equation} \label{eq:exact_energy}
    E_\theta=\frac{\bra{\psi_\theta}\hat{H}\ket{\psi_\theta}}{\braket{\psi_\theta}{\psi_\theta}}=\mathbb{E} \left[E_{l}(\mathbf{x})\right]
\end{equation} where $\mathbb{E}[\cdot]=\mathbb{E}_{p_{\theta}(\mathbf{x})}[\cdot]$ for brevity, and the local energy is $E_{l}(\mathbf{x}) = \frac{\bra{\psi_\theta}\hat{H}\ket{\mathbf{x}}}{\bra{\psi_\theta}\ket{\mathbf{x}}}$, with the quantum chemical Hamiltonian taking the form
\begin{equation}\label{eq:hamiltonian}
    \hat{H} = \sum_{i,j,\sigma}t_{ij}\hat{c}_{i,\sigma}^\dagger \hat{c}_{j,\sigma} + \frac{1}{2} \sum_{i,j,k,l,\sigma,\sigma'}V_{ijkl}\hat{c}_{i,\sigma}^\dagger  \hat{c}_{j,\sigma'}^\dagger \hat{c}_{l,\sigma'} \hat{c}_{k,\sigma}.
\end{equation}
where indices $i,j,k,l$ iterate over the $N_o$ SPOs and $\sigma, \sigma'$ denote spin. The gradient of energy is then given by
\begin{equation} \label{eq:exact_gradient}
    \nabla_\theta E_\theta = 2\Re{ \mathbb{E}\left[ \frac{\partial \ln{\abs{\psi_\theta(\mathbf{x})}}}{\partial \theta} \left[E_{l}(\mathbf{x}) - E_\theta \right] \right] } 
\end{equation},
which allows iterative updates of $\theta$ using gradient descent or more advanced optimization techniques.

For molecular Hamiltonians, previous studies \cite{Choo2020,Liu2024} have shown that estimating equation \eqref{eq:exact_energy} and \eqref{eq:exact_gradient} using conventional MCMC methods, such as the Metropolis–Hastings algorithm, is inefficient. This inefficiency stems from pronounced peaks around the Hartree–Fock (HF) state and nearby excited states \cite{Bytautas2009,Anderson2018}, which lead to excessive resampling of dominant configurations and thus waste computational resources. Another computational challenge related to the molecular Hamiltonian is that the local energy calculation involves a number of terms that grows quartically with the number of orbitals, causing the computation to become increasingly burdensome as the system size grows.

To mitigate the inefficient sampling issue, several techniques have been developed. Autoregressive neural quantum states, combined with improved sampling methods, enable exact sampling from the Born distribution. Alternatively, approaches inspired by SCI bypass stochastic sampling altogether by deterministically identifying important states and approximating their relative contributions during energy evaluation \cite{Li2023,Liu2024,Li2024}.

In this work, we build on the latter, SCI-inspired approach.  There is a core space $\mathcal{V}$, containing a set of unique and dominant configurations, and a target space $\mathcal{U}$, containing a larger compact yet relevant set of important determinants, both of which are updated every $l$ optimization steps.  This periodic update takes as input the current $\mathcal{U}$ as well as a set of walkers generated stochastically from MCMC.  From $\mathcal{U}$, we  generate a more efficient and effective approximation of the local energy $E_l(\mathbf{x}|\mathcal{U})$. Additionally, we develop an improved strategy for sampling configurations $\mathcal{S}$ and assigning importance weights $w$, resulting in more accurate estimates of equation \eqref{eq:exact_energy} and \eqref{eq:exact_gradient}.  These are the critical steps required to improve the accuracy and efficiency of the optimization.

\subsection{Intermittent Target Selection}\label{sec:ITS}

In this subsection, we introduce Intermittent Target Selection (ITS), a method to construct and periodically update a compact yet highly relevant subspace, denoted as $\mathcal{U}$, every $l$ optimization steps. This subspace and its amplitudes, which are calculated at each step, are then used in subsequent optimization subroutines including sample drawing and local energy computations to improve computational efficiency.

The output from ITS is the new target space $\mathcal{U}$ with input as the $\mathcal{U}$ that was selected $l$ steps ago and has been fixed in the last $l$ steps. 
At this point in the algorithm, we have $\psi_\theta(\mathbf{x}_i)$ for all $\ket{\mathbf{x}_i} \in \mathcal{U}$ computed from the last optimization step. 
The first step of ITS is to re-generate a new core space $\mathcal{V}$ by choosing $\abs{\mathcal{V}}$ unique configurations with the highest amplitude moduli from both $\mathcal{U}$ as well as the $\abs{\mathcal{V}}$ samples generated from a parallel MCMC track, which gives access to stochastic samples from the current wave-function - i.e.
\begin{equation}
\mathcal{V} \leftarrow \underset{\ket{\mathbf{x}_i} \in \mathcal{MC} \cup \mathcal{U}}{\text{argtop}(\abs{\mathcal{V}})} \left[|\psi_\theta(\mathbf{x}_i)|\right]
\end{equation}
The cost of this step (on top of running the parallel MCMC track) is computationally inexpensive, requiring no new NNBF evaluations and only a sorting cost of $O((\abs{\mathcal{U}}+\abs{\mathcal{V}}) \log(\abs{\mathcal{U}}+\abs{\mathcal{V}})$. Now, as in the language of SCI, the core space $\mathcal{V}$ is connected to other configurations via non-zero Hamiltonian matrix elements, collectively forming the connected space $\mathcal{C}$. 
As the system size increases, the size of $\mathcal{C}$ grows quartically with a fixed core space size $\abs{\mathcal{V}}$, making it computationally prohibitive to include all of these connected configurations in every optimization step. However, since the ground state is typically dominated by a small fraction of these configurations, a compact (reduced by $l$ times) subspace $\mathcal{U}$ can be identified by selecting the $\abs{\mathcal{C}}/l$ unique configurations with largest amplitude moduli (with respect to the current wave-function) from $\mathcal{V} \cup \mathcal{C}$ - i.e.
\begin{equation}
\mathcal{U} \leftarrow \underset{\ket{\mathbf{x}_i} \in \mathcal{V} \cup \mathcal{C}}{\text{argtop}\left(\frac{\abs{\mathcal{C}}}{l}\right)} \left[|\psi_\theta(\mathbf{x}_i)|\right]
\end{equation}
The update of $\mathcal{U}$ is the dominant cost of the cycle, requiring $\abs{\mathcal{C}}\sim \abs{\mathcal{V}}N_e^2(N_o-N_e)^2$ NNBF evaluations (plus an additional sorting cost).

This strategy effectively amortizes the high cost of exploring the full connected space. The average number of amplitude evaluations is reduced by a factor of approximately $l$ compared to a naive approach that evaluates the entire space $\mathcal{C}$ at every step. The interval $l$ is chosen to align this amortized cost with the per-step cost of amplitude computations within $\mathcal{U}$, which requires $\abs{\mathcal{C}}/l$ NNBF evaluations; we typically use $l \sim (N_o-N_e)$. To manage the computational cost of the MCMC evolution, the number of MCMC walkers, $N_w$, is typically set to equal the core space size ($N_w = \abs{\mathcal{V}}$), and each walker performs $N_e$ proposed hopping moves per optimization step.

The effectiveness of ITS hinges on the assumption that the wavefunction parameters, $\theta$, evolve slowly enough that the changes over $l$ steps are minor. This ensures that the subspace $\mathcal{U}$ selected at the beginning of a cycle remains a good approximation of the most dominant configurations for the entire interval. Such slow parameter evolution is typically achieved by using a small learning rate and is most prominent as the optimization converges and the gradients naturally diminish.

\subsection{Streamlined Local Energy Calculations}\label{sec:truncated_local_energy}

Building upon the Intermittent Target Selection (ITS) strategy detailed in Section~\ref{sec:ITS}, this subsection introduces an approach to accelerate the calculation of the local energy—identified as one of the primary computational bottlenecks in NQS-based quantum chemistry. Computing $E_{l}(\mathbf{x}_i)=\sum_{\ket{\mathbf{x}_j}}H_{ij}\frac{\psi(\mathbf{x}_i)}{\psi(\mathbf{x}_j)}$ requires evaluating ansatz amplitudes for all $\ket{\mathbf{x}_j}$ connected to $\ket{\mathbf{x}_i}$ through nonzero $H_{ij}$. For second-quantized molecular Hamiltonians, the number of such terms grows quartically with system size, making this step computationally expensive.

To mitigate this cost, we introduce a truncated approximation of the local energy by leveraging the selectively important subspace $\mathcal{U}$ from Section \ref{sec:ITS}:
\begin{equation}\label{eq:truncated_el}
E_{l}(\mathbf{x}_i)=E_l(\mathbf{x}_i|\mathcal{U})=\sum_{\ket{\mathbf{x}_j}\in\mathcal{U}}H_{ij}\frac{\psi(\mathbf{x}_j)}{\psi(\mathbf{x}_i)}.
\end{equation}
Because the amplitudes for configurations in $\mathcal{U}$ are already computed, the evaluation of this truncated local energy avoids repeated, costly wavefunction evaluations. Storing $\mathcal{U}$ in lexicographical order further enables efficient amplitude retrieval through $O(\log(\abs{\mathcal{U}}))$ lookups.

Our approach for computing the local energy shares conceptual similarities with other recent methods \cite{Wu2023,Li2024,Malyshev2024}, where the local energy sum is restricted to the sample set $\mathcal{S}$ generated by their sampling schemes. However, a key distinction is that $\mathcal{U}$ in our method is both substantially larger than $\mathcal{S}$ and constructed for importance. This strategy ensures that the local energy calculation retains more significant contributions, leading to a more accurate approximation and improved training performance at a comparable computational cost. A direct comparison of these strategies is presented in Section \ref{sec:local_energy_strategy_comparison}.

\subsection{Gumbel top-k trick}\label{sec:gumbel}
In this section, we introduce a new sampling method that provides better approximations of the energy and its gradients than the estimates given by the fixed-size selected configuration (FSSC) scheme \cite{Liu2024}, while maintaining the same time complexity. Although Ref.~\onlinecite{Liu2024} has shown that, for a given batch size, the FSSC scheme outperforms the MCMC scheme by capturing the most significant distinct configurations and avoiding sequential, redundant stochastic sampling, there is still room for improvement.

\begin{figure}[htbp!]
    \centering
    \includegraphics[width=\linewidth]{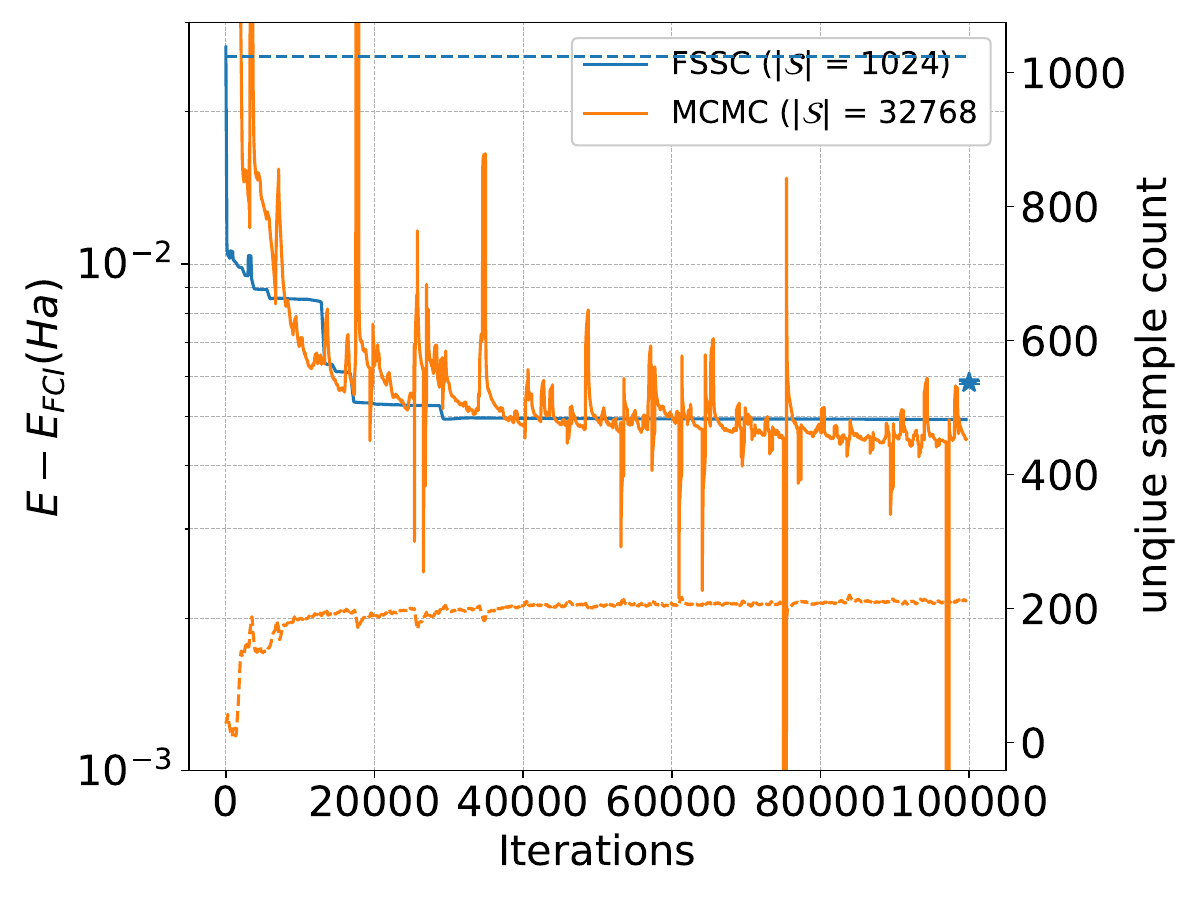}
    \caption{Comparison of the energy optimization curve and the number of unique configurations sampled per iteration between the FSSC \cite{Liu2024} ($\abs{\mathcal{S}}=1024$) and MCMC ($\abs{\mathcal{S}}=32768$) schemes for Li$_2$O, using the STO-3G basis set and canonical HF orbitals, where $\abs{\mathcal{S}}$ denotes the size of sample set. Solid lines represent the objective energy optimization curve, while dashed lines indicate the number of unique configurations sampled per iteration. The blue star marks the post-training MCMC inference energy for the FSSC scheme. A moving average window of 400 is applied for improved readability.}
    \label{fig:MCMCvsFSSC}
\end{figure}

A key observation is that borderline configurations—those whose amplitude magnitudes are just below the smallest amplitude in the selected sample—may contribute nontrivially to the energy and gradient evaluations. As optimization progresses, a small batch size combined with the deterministic nature of the FSSC scheme may prevent these borderline configurations from being selected, potentially introducing bias. To investigate this, we perform a simple test on the Li$_2$O molecule in the STO-3G basis set comparing the FSSC and standard MCMC schemes. As shown in Fig.~\ref{fig:MCMCvsFSSC}, the MCMC scheme achieves a lower variational energy with fewer unique configurations, even though it requires a larger total batch size and therefore is slower. This observation on one hand reinforces the inefficiency of standard MCMC sampling for second-quantized molecular simulations, while on the other hand suggesting that stochastic estimation—when paired with an optimization method leveraging the entire optimization history for parameter updates—could possibly yield estimates that are less biased than purely deterministic methods with the same number of unique samples.

This finding motivates us to enhance the FSSC scheme by incorporating stochasticity while preserving its unique sample selection feature, i.e., sampling without replacement (SWOR). To achieve this, we employ the Gumbel top-$k$ trick \cite{Vieira2014,Kool2019}, a powerful SWOR technique that extends the Gumbel-max trick. By adding independently sampled Gumbel noise to the (unnormalized) log-probabilities of categories and selecting the top-$k$ perturbed values, one can sample from the categorical distribution without replacement. Importantly, Gumbel noise sampling is highly parallelizable, making it suitable for efficient GPU implementation, and it yields unbiased estimates when properly weighted \cite{Kool2019,Vieira2014}.

A natural candidate for applying the Gumbel top-$k$ trick is the target space $\mathcal{U}$ from Section \ref{sec:ITS}. Unlike the original FSSC scheme, which relies on the dominance of the core space $\mathcal{V}$, using the Gumbel top-$k$ trick allows us to unbiasedly represent $\mathcal{U}$, which is more representative than $\mathcal{V}$ alone, as it is constructed for importance and grows extensively with system size when $\abs{\mathcal{V}}$ is fixed. By introducing Gumbel noise, previously excluded borderline configurations can be sampled, improving the approximation of both energy and gradients. Moreover, the computational overhead is minimal since Gumbel noise generation is inexpensive.

To formalize this approach, we describe how the Gumbel top-$k$ trick is used to form the sample $\mathcal{S}$ and assign importance weights $w_i$ for constructing more unbiased estimates of equation \eqref{eq:exact_energy} and \eqref{eq:exact_gradient}. 
The procedure begins with a normalized probability distribution over the target space $\mathcal{U}$, derived from the precomputed amplitudes $\psi_\theta(\mathbf{x}_i) \ \forall\ket{\mathbf{x}_i}\in\mathcal{U}$:
\begin{equation}\label{eq:target_space_prob}
     p_i = \frac{\psi^2_\theta(\mathbf{x}_i)}{\sum_{\ket{\mathbf{x}_j}\in\mathcal{U}}\psi^2_\theta(\mathbf{x}_j)}, \quad \forall \ket{\mathbf{x}_i} \in \mathcal{U}.
\end{equation}
Samples are drawn from $\mathcal{U}$ by first perturbing this log probability with independently and identically distributed Gumbel noise:
\begin{equation} 
g_i \overset{i.i.d.}{\sim} \text{Gumbel}(0,1) \quad \text{with pdf} \quad f(x)=e^{-(x+e^{-x})}, 
\end{equation}
and then selecting the top-$k$ configurations with perturbed log-probabilities:
\begin{equation} 
\mathcal{S} = \underset{\ket{\mathbf{x}_i}\in\mathcal{U}}{\text{argtop}k}[G_i=\log{p}_i+g_i].
\end{equation}

To construct an unbiased estimator over the target space $\mathcal{U}$, we follow the Horvitz-Thompson estimator by assigning each sample an importance weight, $w_i$, defined as the ratio of its original probability $p_i$ to its inclusion probability $q_i(\kappa)$\cite{Kool2019,Vieira2014}. The inclusion probability $q_i(\kappa)$—the probability of configuration $i$ being selected—is determined by an empirical threshold $\kappa$ set by the $(k+1)$-th largest perturbed log-probability, $G_i$. The formulas for the inclusion probability and the final weight are:
\begin{align}
    q_i(\kappa) &= P(G_i>\kappa) = 1 - e^{-e^{(\log{p}_i-\kappa)}} \label{eq:inclusion_probability} \\
    w_i(\kappa) &= \frac{p_i}{q_i(\kappa)} = \frac{p_i}{1 - e^{-e^{(\log{p}_i-\kappa)}}}\label{eq:gumbel_weight}
\end{align}
Importance weights are typically renormalized over sample $\mathcal{S}$, i.e. $w_i\leftarrow w_i/\sum_{j\in\mathcal{S}}w_j$, to reduce variance in practice \cite{Kool2019}, albeit at the cost of introducing bias. The effects of the Gumbel noise, the use of inclusion probabilities for weighting, and weight renormalization are investigated in detail in Section \ref{sec:effect_gumbel_feature}.

Using these weights and the Gumbel-top-$k$-selected samples $\mathcal{S}$, we construct improved estimates for the energy and gradient:
\begin{equation}  \label{eq:gumbel_energy}
E_\theta\approx \sum_{\ket{\mathbf{x}_i}\in\mathcal{S}} w_i(\kappa) E_{l}(\mathbf{x}_i|\mathcal{U})
\end{equation}
and
\begin{equation} \label{eq:gumbel_gradient}
\nabla_\theta E_\theta \approx 2\Re{  \sum_{\ket{\mathbf{x}_i}\in\mathcal{S}} w_i(\kappa)\left[E_{l}(\mathbf{x}_i|\mathcal{U}) - E_\theta  \right]\frac{\partial \ln{\abs{\psi_\theta(\mathbf{x}_i)}}}{\partial \theta}  }.
\end{equation}
where $E_{l}(\mathbf{x}_i|\mathcal{U})$ is equation \eqref{eq:truncated_el}. While equation \eqref{eq:gumbel_energy} and \eqref{eq:gumbel_gradient} are not fully unbiased estimators of equation \eqref{eq:exact_energy} and \eqref{eq:exact_gradient}, they significantly reduce bias compared to the deterministic FSSC approach.

\subsection{Encode physical knowledge}\label{sec:encode_physics}

\subsubsection{Enforce Spin-flip symmetry}\label{sec:spin_flip_symmetry}
In this work, we propose a general approach to enforce the spin-flip symmetry on top of the NNBF ansatz. We define the spin-flip operator $\hat{F}$ as a transformation that exchanges the spin components of a configuration: 
$\hat{F}:\ket{x_i^{1\uparrow},\cdots,x_i^{N_o\uparrow},x_i^{1\downarrow},\cdots,x_i^{N_o\downarrow}} \mapsto \ket{x_i^{1\downarrow},\cdots,x_i^{N_o\downarrow},x_i^{1\uparrow},\cdots,x_i^{N_o\uparrow}}$. For eigenstates $\ket{\psi}$ with total spin zero, this symmetry implies $\hat{F}\ket{\psi} = \pm\ket{\psi}$, and thus the amplitudes of spin-flip-equivalent configurations, $\ket{\mathbf{x}_i}$ and $\ket{\hat{F}\mathbf{x}_i}$, satisfy $\psi(\hat{F}\mathbf{x}_i) = \pm \psi(\mathbf{x}_i)$. This condition holds in second-quantized molecular systems where spin orbitals are constructed with spin-independent spatial components, as in restricted HF methods.

We impose this symmetry by defining the spin-flip-symmetric wavefunction as:
\begin{equation}
    \psi_{SFS,\theta}(\mathbf{x}) = \psi_\theta(\mathbf{x}) \pm \psi_\theta(\hat{F}\mathbf{x}).
\end{equation}
This construction ensures both $\hat{F}\ket{\psi_{SFS,\theta}} = \pm \ket{\psi_{SFS,\theta}}$ and $\psi_{SFS,\theta}(\hat{F}\mathbf{x}_i) = \pm \psi_{SFS,\theta}(\mathbf{x}_i)$. Unlike the approach in Ref. \cite{Barrett2022}, which enforces a weaker form of the symmetry ($\abs{\psi(\hat{F}\mathbf{x}_i)}=\abs{\psi(\mathbf{x}_i)}$) through specialized preprocessing of subnetwork inputs and postprocessing of outputs, our method is more general. It requires only the addition of a single equation atop any wavefunction ansatz, ensuring full spin-flip symmetry while naturally preserving the relative sign consistency between spin-flip-equivalent configurations.

\subsubsection{Spin-flip-symmetry-aware strategies}\label{sec:sfs_trick}

When a wavefunction ansatz is constructed with spin-flip symmetry, as described in Section~\ref{sec:spin_flip_symmetry}, several strategies can be leveraged to enhance the training pipeline.

The process of updating the subspace $\mathcal{U}$ (see Section \ref{sec:ITS}) can explicitly incorporate spin-flip symmetry. For any pair of spin-flip-equivalent configurations (or for a single configuration if it is self-spin-flip-equivalent), only one unique representative is considered to be selected into $\mathcal{U}$, determined by a predefined rule. This choice is justified because both configurations in such a pair share an identical amplitude magnitude. This initial filtering of $\mathcal{U}$ to remove spin-based redundancy provides several downstream advantages.

First, this efficiency propagates to the construction of the core and connected spaces. The two sets of configurations connected to any spin-flip-equivalent pair are themselves spin-flip equivalent. Therefore, if the core space $\mathcal{V}$ is selected from the now redundancy-free subspace $\mathcal{U}$, then the expansion of connected configurations from this new $\mathcal{V}$ will also avoid such spin-flip redundancies.

Second, during sample generation from $\mathcal{U}$ via the Gumbel top-$k$ trick, the treatment of non-spin-flip-symmetric configurations can be adjusted. Consider a configuration $\ket{\mathbf{x}_i} \in \mathcal{U}$ that is part of a non-symmetric pair. Its partner, $\hat{F}\ket{\mathbf{x}_i}$, is not stored in $\mathcal{U}$. However, both configurations have the same probability ($p(\mathbf{x}_i) = p(\hat{F}\mathbf{x}_i)$) and the same local energy. To account for the excluded partner, we effectively transfer its importance weight to the representative configuration stored in $\mathcal{U}$ by doubling its sampling probability:
\begin{equation}
    p_{\text{SFS},i} =
    \begin{cases}
    2p_i, & \text{if } \ket{\mathbf{x}_i} \in \mathcal{U} \text{ is non-spin-flip-symmetric}, \\
    p_i, & \text{if } \ket{\mathbf{x}_i} \in \mathcal{U} \text{ is spin-flip-symmetric},
    \end{cases}
\end{equation}
where $p_i$ refers to the probabilities derived from Eq.~\eqref{eq:target_space_prob}. These modified probabilities are then used in the Gumbel top-$k$ sampling and can be renormalized as needed.

Third, this symmetry awareness extends to the evaluation of the truncated local energy (Eq.~\eqref{eq:truncated_el}). Although $\mathcal{U}$ stores amplitudes $\psi_\theta(\mathbf{x}_i)$ only for unique representatives, the amplitude of a spin-flipped partner, $\psi_\theta(\hat{F}\mathbf{x}_i)$, is also implicitly known through the symmetry relation $\psi_\theta(\hat{F}\mathbf{x}_i) = \pm\psi_\theta(\mathbf{x}_i)$. Thus, when computing local energies, all connected configurations are first transformed by the predefined rule. The amplitude information of these transformed connected configurations is then retrieved from $\mathcal{U}$, and the spin-flip-symmetry phase will be applied to the retrieved amplitude if the connected configuration was indeed spin-flipped. This effectively allows the local energy calculation to leverage amplitude information from an expanded set of configurations.

These modifications effectively increase the utilized sample size and information content—often doubling it, since true spin-flip-symmetric configurations are typically far less numerous than their non-symmetric counterparts—all without compromising the efficiency of the truncated local energy strategy.

\subsubsection{Orbital occupation}\label{sec:orb_env}

We introduce a trainable discrete orbital envelope designed to capture general occupation patterns of molecular orbitals. In quantum chemistry, it is well established that lower-energy orbitals have higher occupation probabilities, with core electrons typically occupying the lowest-energy orbitals. Consequently, configurations further from the HF reference generally have lower probabilities.

To incorporate this prior knowledge, we first define an ordered occupied-position representation of the configuration bitstring as $\mathbf{y}^{\uparrow/\downarrow}(\mathbf{x})=vector(\{i|x^{i\uparrow/\downarrow}=1\})$, which lists the indices of occupied spin up/down orbitals in ascending order. Using this representation, we propose the following trainable discrete orbital envelope:
\begin{align}\label{eq:orbital_envelope}
\pi_{\boldsymbol{\alpha}}(\mathbf{x}) = \exp\Biggl(-\sum_{i=1}^{N_e/2} \left[  |\alpha_i(y^{\uparrow}_i - i)| +  |\alpha_i(y^{\downarrow}_i - i)| \right] \Biggr),
\end{align}
where $\boldsymbol{\alpha}=\{\alpha_i\}$ is a set of trainable parameters.
The term $|y^{\sigma}_i - i|$ measures the displacement of the $i$-th electron of spin $\sigma\in\{\uparrow,\downarrow\}$ from its reference orbital index $i$. Each parameter $\alpha_i$ learns a penalty for this displacement, effectively suppressing configurations where electrons are excited into higher-energy orbitals and favoring lower-indexed electrons for lower-indexed orbitals. To respect spin-flip symmetry, both spin channels share the same set of envelope parameters $\boldsymbol{\alpha}$. A visualization and analysis of the learned orbital envelope parameters after training are provided in Appendix \ref{appx:learned_orbital_envelope}.

\subsubsection{Selection of orbitals}\label{sec:orb_rot}

In the framework of SCI, the choice of single-particle orbitals (molecular orbitals) significantly affects the compactness of the wavefunction and, consequently, the convergence of the simulation with respect to the number of determinants. Selecting orbitals that achieve a given accuracy with the fewest configurations is therefore crucial.

A common choice is the set of natural orbitals \cite{Lowdin1955}, defined as the eigenstates of the 1-RDM derived from other wavefunctions, such as those from HF, MP2, CISD, or CCSD. Expansions based on natural orbitals generally converge more rapidly than those using canonical HF orbitals.

For all-electron calculations, the selection of single-particle orbitals does not influence the exact ground-state energy. Hence, unless otherwise specified, we use CCSD natural orbitals in all-electron calculations to promote faster convergence.


\section{Results}
We first evaluate the performance of NNBF combined with the algorithmic enhancements proposed in Section \ref{sec:ITS}, \ref{sec:truncated_local_energy}, \ref{sec:gumbel}, and \ref{sec:encode_physics} on various molecules using the STO-3G basis set as well as on paradigmatic strongly correlated systems: the dissociation curve of H$_2$O utilizing the cc-pVDZ basis set, and frozen-core and all-electron N$_2$ molecule calculations also with the cc-pVDZ basis set at representative bond lengths. 

To demonstrate the improvements from these enhancements, we present three distinct analyses. First,we conduct a thorough ablation study by cumulatively adding the proposed techniques starting from our previous algorithm \cite{Liu2024}. Second, we perform a focused study on the Gumbel top-$k$ selection scheme described in Section~\ref{sec:gumbel}, examining the specific roles of the Gumbel noise, the use of inclusion probabilities, and weight renormalization. Finally, we directly compare our proposed truncated local energy strategy (Section~\ref{sec:truncated_local_energy}) against an approach commonly employed in the community \cite{Wu2023,Li2024,Malyshev2024}.

The relationship between the representational capacity of NNBF and the inverse participation ratio (IPR) of the quantum state is also investigated. Specific details regarding the (default) neural network architectures, hyperparameters, training protocols, and the post-training MCMC inference procedure are provided in Appendix~\ref{appx:experimental_setup}.

\subsection{Benchmarks}
\subsubsection{Ground state energy for various molecules}\label{sec:GSE_mol}
We assess the performance of our enhanced NNBF algorithms by first comparing calculated molecular ground-state energies against those from established CCSD and CCSD(T) baselines, as well as results from other NQS methods. These calculations utilize molecular geometries sourced from \textit{PubChem} \cite{pubchem}, which is also provided in Table \ref{tab:geometries_consolidated} in Appendix~\ref{appx:GSE_mol_geometry} for reference and convenience, and strictly adhere to the computational protocols described in Section~\ref{sec:ITS}, Section~\ref{sec:truncated_local_energy}, Section~\ref{sec:gumbel}, and Section~\ref{sec:encode_physics}. The results, summarized in Table~\ref{tab:molecule_energy}, demonstrate that NNBF employed with the improved algorithms not only generally outperforms conventional CCSD methods and achieves energies comparable or superior to CCSD(T) for many systems, but also consistently yields lower energies than other existing NQS approaches, particularly excelling for larger molecular systems.

\begin{table*}[htbp!] 
  \centering
  
  \begin{tabular}{lllllll}
    \toprule
    Molecule & $\abs{\mathcal{H}}$ & CCSD & CCSD(T) & FCI & Best NQS & NNBF\\ 
    \midrule
    N$_2$ & $1.44\times10^{4}$ &  -107.656080 & -107.657850 &-107.660206 & -107.6602\textsuperscript{\cite{Shang2023}} & -107.660218(67) \\
    CH$_4$ & $1.59\times10^{4}$ &  -39.806022 & -39.806164 & -39.806259 & -39.8062\textsuperscript{\cite{Shang2023}} & -39.806258(22)\\
    LiF & $4.41\times10^{4}$ & -105.159235 & -105.166274 & -105.166172 & -105.1661\textsuperscript{\cite{Shang2023}} & -105.166169(18)\\
    CH$_2$O & $2.45\times10^{5}$ &  -112.498566 & -112.500465 & -112.501253 & -112.500944\textsuperscript{\cite{Zhao2023}} & -112.501201(9)\\
    LiCl & $1.00\times10^{6}$ & -460.847579 & -460.849980 & -460.849618 & -460.8496\textsuperscript{\cite{Barrett2022}} & -460.849614(10)\\
    CH$_4$O & $4.01\times10^{6}$ & -113.665485 & -113.666214 & -113.666485 & -113.665485\textsuperscript{\cite{Zhao2023}} & -113.666416(26)\\
    Li$_2$O & $4.14\times10^{7}$ &  -87.885514 & -87.893089 & -87.892693 & -87.8922\textsuperscript{\cite{Shang2023}} & -87.892662(18)\\
    C$_2$H$_4$O & $2.54\times10^{9}$ & -151.120474 & -151.122748 & -151.123570 & -151.12153\textsuperscript{\cite{Shang2023}} & -151.123357(28)\\
    C$_2$H$_4$O$_2$ & $5.41\times10^{11}$ & -225.050896 & -225.057823 & \makecell[c]{-} & -225.0429767\textsuperscript{\cite{Zhao2023}} & -225.058589(40)\\
    \bottomrule
  \end{tabular}
  
  \caption{
  A benchmark of ground-state energies from our NNBF method against conventional quantum chemistry methods (CCSD, CCSD(T), FCI) as well as the best published NQS results, with footnotes indicating the respective methods (excluding our previous work \cite{Liu2024}).  $|\mathcal{H}|$ is the size of the total Hilbert space, comprising all configurations that conserve both particle number and total spin projection ($S_z$). The reported NNBF energy is obtained via the following protocol: five independent training runs were performed (settings in Appendix \ref{appx:experimental_setup}), and a post-training MCMC inference was used to estimate the energy of each. The single model with the lowest of these five energies was selected. A final, separate MCMC inference was then conducted on this best model to obtain the unbiased estimate reported in the table.
  } 
  \label{tab:molecule_energy}
\end{table*}

\subsubsection{All electron \texorpdfstring{H$_2$O}{H2O} dissociation curve}
\begin{figure}[htbp!]
    \centering
    \includegraphics[width=\linewidth]{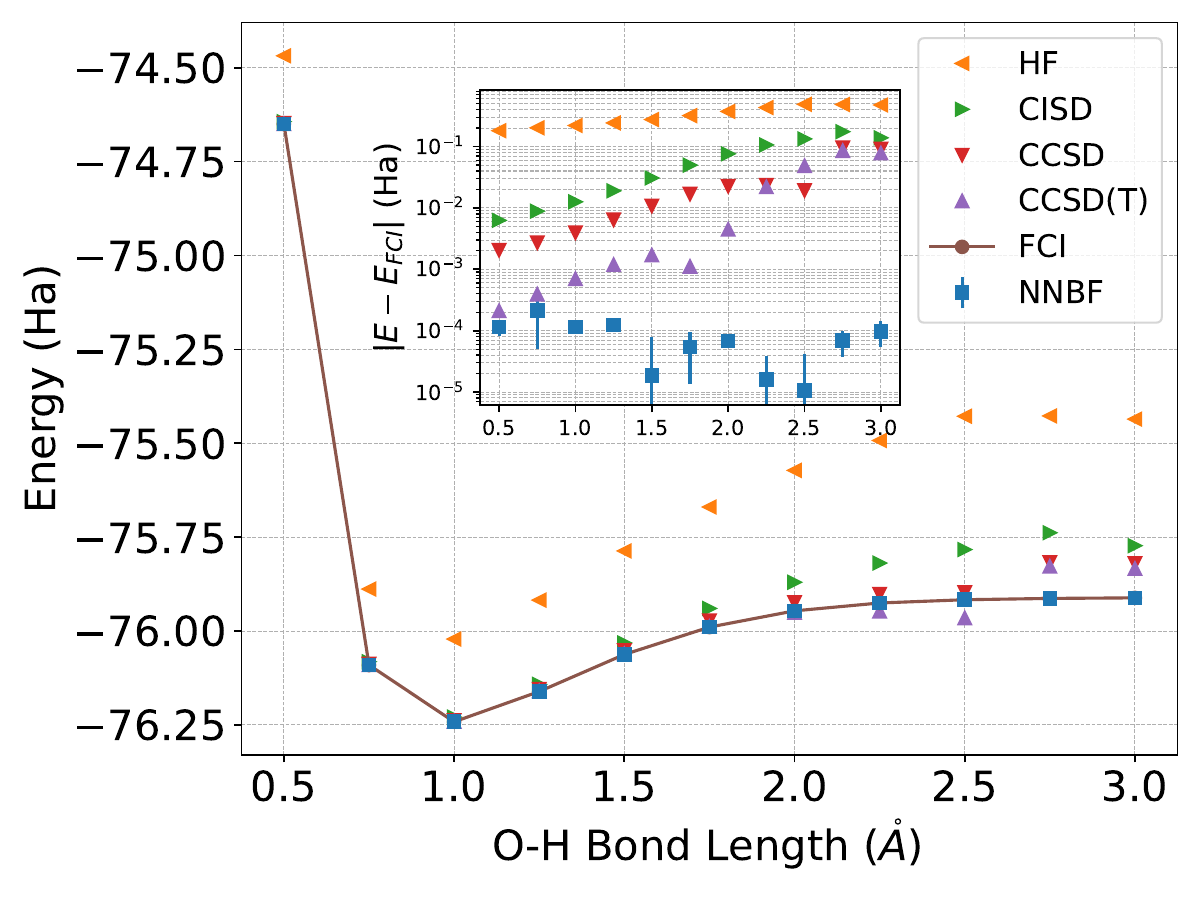}
    \caption{
    Dissociation curve of the H$_2$O molecule, calculated using a cc-pVDZ basis set with the bond angle fixed at $104.5^{\circ}$. The plot compares the performance of our NNBF method against standard quantum chemistry approaches (HF, CISD, CCSD, CCSD(T)) and the exact FCI energy, which serves as the ground-truth reference. The NNBF results were obtained from a single training run per bond length with parameters $|\mathcal{S}|=8192$, $|\mathcal{V}|=2048$, and a $(D,L,h)=(1,2,512)$. Each reported NNBF energy is the result of a single post-training MCMC inference.
    }
    \label{fig:H2O}
\end{figure}

To investigate the ability of the NNBF method to describe strong quantum correlations, we computed the dissociation curve for an all-electron H$_2$O molecule using the cc-pVDZ basis set, with the bond angle held fixed at $104.5^{\circ}$. As illustrated in Figure~\ref{fig:H2O}, the resulting NNBF energies are in excellent agreement with the FCI benchmark, achieving a mean absolute error of only 0.08 mHa across the entire curve. Notably, NNBF outperforms conventional quantum chemistry approaches for both near-equilibrium geometries and stretched bond lengths. The latter is a region where the gold-standard CCSD(T) method is known to falter due to the increasing importance of static correlation at large bond separations. This result demonstrates the proficiency of NNBF in accurately capturing both static and dynamic electron correlations.

\subsubsection{All-electron and frozen-core \texorpdfstring{N$_2$}{N2} calculations}
To further evaluate our method against other state-of-the-art ab-initio techniques, we calculated the ground-state energy of frozen-core N$_2$ with the cc-pVDZ basis set. The results, presented in Table~\ref{tab:frozen_core_N2}, show that when using canonical Hartree-Fock orbitals, NNBF achieves superior variational energies compared to prominent SCI methods such as SCHI \cite{Sharma2017} and ASCI \cite{Tubman2018}. 
After applying an orthonormal rotation to the orbitals—an operation that leaves the exact ground-state energy invariant—the NNBF energy further improves 
and is in excellent agreement with FCIQMC \cite{Cleland2012} as well as the perturbatively corrected values from SCHI and ASCI. This performance underscores the NNBF's capacity for accurately modeling strongly correlated systems. In contrast, other NQS methods have reported energy differences of several mHa for this system \cite{Li2024}, further emphasizing the robustness of the NNBF ansatz and the efficacy of our algorithmic enhancements.

Moreover, we performed all-electron calculations for the N$_2$ molecule at selected bond lengths using the cc-pVDZ basis set. The data in Table~\ref{tab:all_electron_N2} demonstrate that NNBF not only surpasses the accuracy of conventional quantum chemistry methods up to CCSDTQ but also competes effectively with state-of-the-art Density Matrix Renormalization Group (DMRG) calculations using a bond-dimension of $m=2000$ \cite{Chan2004}. These findings indicate that NNBF is among the first NQS approaches capable of tackling such complex, strongly correlated chemical systems with high accuracy. This success opens new avenues for optimizing large-scale NQS wavefunctions more efficiently and effectively in challenging chemical systems.

\begin{table}[htbp!]
  \centering
  \begin{tabular}{lll}
    \toprule
    \makecell{Method\\(Parameter)} & \makecell{Variational \\ energy} & \makecell{Total \\ energy} \\
    
    \midrule
    HF  & -108.954125 & \makecell[c]{-} \\
    CISD & -109.242435 & \makecell[c]{-} \\
    CCSD  & -109.263394 & \makecell[c]{-} \\
    CCSD(T) & - & -109.275256 \\
    \makecell[l]{SHCI-HF\\(N$_{dets}$=37593)\textsuperscript{\cite{Sharma2017}}}
      & -109.2692 & -109.2769 \\
    \makecell[l]{ASCI-HF\\(N$_{dets}$=10000)\textsuperscript{\cite{Tubman2018}}}&-109.26419&-109.27687\\
    \makecell[l]{ASCI-HF\\(N$_{dets}$=30000)\textsuperscript{\cite{Tubman2018}}}&-109.26936&-109.27691\\
    \makecell[l]{ASCI-HF\\(N$_{dets}$=100000)\textsuperscript{\cite{Tubman2018}}}&-109.27335&-109.27698\\
    \makecell[l]{ASCI-NOR\\(N$_{dets}$=10000)\textsuperscript{\cite{Tubman2018}}}&-109.26837&-109.27708\\
    \makecell[l]{ASCI-NOR\\(N$_{dets}$=100000)\textsuperscript{\cite{Tubman2018}}}&-109.27522&-109.27699\\
    \makecell[l]{ASCI-NOR\\(N$_{dets}$=300000)\textsuperscript{\cite{Tubman2018}}}&-109.27638&-109.27699\\
    FCIQMC - HF\textsuperscript{\cite{Cleland2012}}&\makecell[c]{-}&-109.2767(1)\\
    \makecell[l]{NNBF-HF\\($\abs{\mathcal{S}}$=16384,\\$\abs{\mathcal{V}}$=4096)}&-109.276642(36)&\makecell[c]{-}\\
    \makecell[l]{NNBF-NOCCSD\\($\abs{\mathcal{S}}$=16384,\\$\abs{\mathcal{V}}$=4096)}&-109.276911(54)&\makecell[c]{-}\\ 
    \bottomrule
  \end{tabular}
  
  \caption{
  Ground-state energy benchmark for the frozen-core N$_2$ molecule at a bond length of 1.0977 $\mathring{A}$ with the cc-pVDZ basis set.
  Results from our NNBF method are compared against conventional quantum chemistry (HF, CISD, CCSD, CCSD(T)) and state-of-the-art techniques, including Semistochastic Heat-bath CI (SHCI) \cite{Sharma2017}, Adaptive Sampling CI (ASCI) \cite{Tubman2018}, and Full Configuration Interaction Quantum Monte Carlo (FCIQMC) \cite{Cleland2012}.
  The labels following each method name (e.g., ASCI-NOR) denote the molecular orbitals used. \textbf{-HF} indicates canonical Hartree-Fock orbitals, while \textbf{-NOR} refers to natural orbitals generated from a preliminary calculation, such as the growth phase of the ASCI method.
  For SHCI and ASCI, the ``Variational'' energy is from an exact diagonalization of a determinant space of size $N_{\text{dets}}$, while the ``Total'' energy includes a second-order perturbative correction. The FCIQMC energy is non-variational. The NNBF results are reported for two separate calculations: one using canonical RHF orbitals and another using natural orbitals obtained from a CCSD calculation. Each NNBF energy is variational, determined from a single post-training MCMC inference with a $(D,L,h)=(1,2,512)$ network architecture.
  } 
  \label{tab:frozen_core_N2}
\end{table}

\begin{table}[htbp!]
  \centering
  \begin{tabular}{cccc}
    \toprule
    \makecell{Method\\(Parameter)} & 2.118 Bohr & 2.7 Bohr & 3.6 Bohr \\
    
    \midrule
    RHF & -108.949378 & -108.833687 & -108.767549 \\
    CCSD & -109.267626 & -109.131491 & -108.975885 \\
    CCSD(T) & -109.28030 & -109.150645 & -108.982836 \\
    CCSDT & -109.280323 & -109.156703 & -108.990518 \\
    CCSDTQ & -109.281943 & -109.162264 & -108.993736 \\
    \makecell{DMRG\\(m=1000)} & -109.281878 & -109.163087 & -108.997549 \\
    \makecell{DMRG\\(m=2000)} & -109.282088 & -109.163467 & -108.997939 \\
    \makecell{DMRG\\(m=4000)} & -109.282157 & -109.163572 & -108.998052 \\
    \makecell{NNBF\\($\abs{\mathcal{S}}$=65536 \\ $\abs{\mathcal{V}}$=16384)} & -109.282036(48) & -109.163438(67) & 
    -108.997789(142)\\ 

    \bottomrule
  \end{tabular}
  
  \caption{
  Ground-state energy benchmark for the all-electron N$_2$ molecule (cc-pVDZ basis set) at three bond lengths: 2.118, 2.7, and 3.6 Bohr. Results from our NNBF method are compared against conventional quantum chemistry methods (HF, CCSD, CCSD(T), CCSDT, and CCSDTQ) and state-of-the-art Density Matrix Renormalization Group (DMRG) calculations \cite{Chan2004}. The DMRG calculations were performed using canonical UHF orbitals with several bond dimensions ($m=1000, 2000, \text{and } 4000$). The NNBF calculations, in contrast, used CCSD natural orbitals. Each reported NNBF energy is variational, obtained from a single post-training MCMC inference with a $(D,L,h)=(1,2,512)$ network architecture.
  } 
  \label{tab:all_electron_N2}
\end{table}


\subsection{Ablation study}\label{sec:ablation_study}
\subsubsection{Cumulative Feature Addition}
\begin{figure*}[htbp!]
    \centering
    \includegraphics[width=\linewidth]{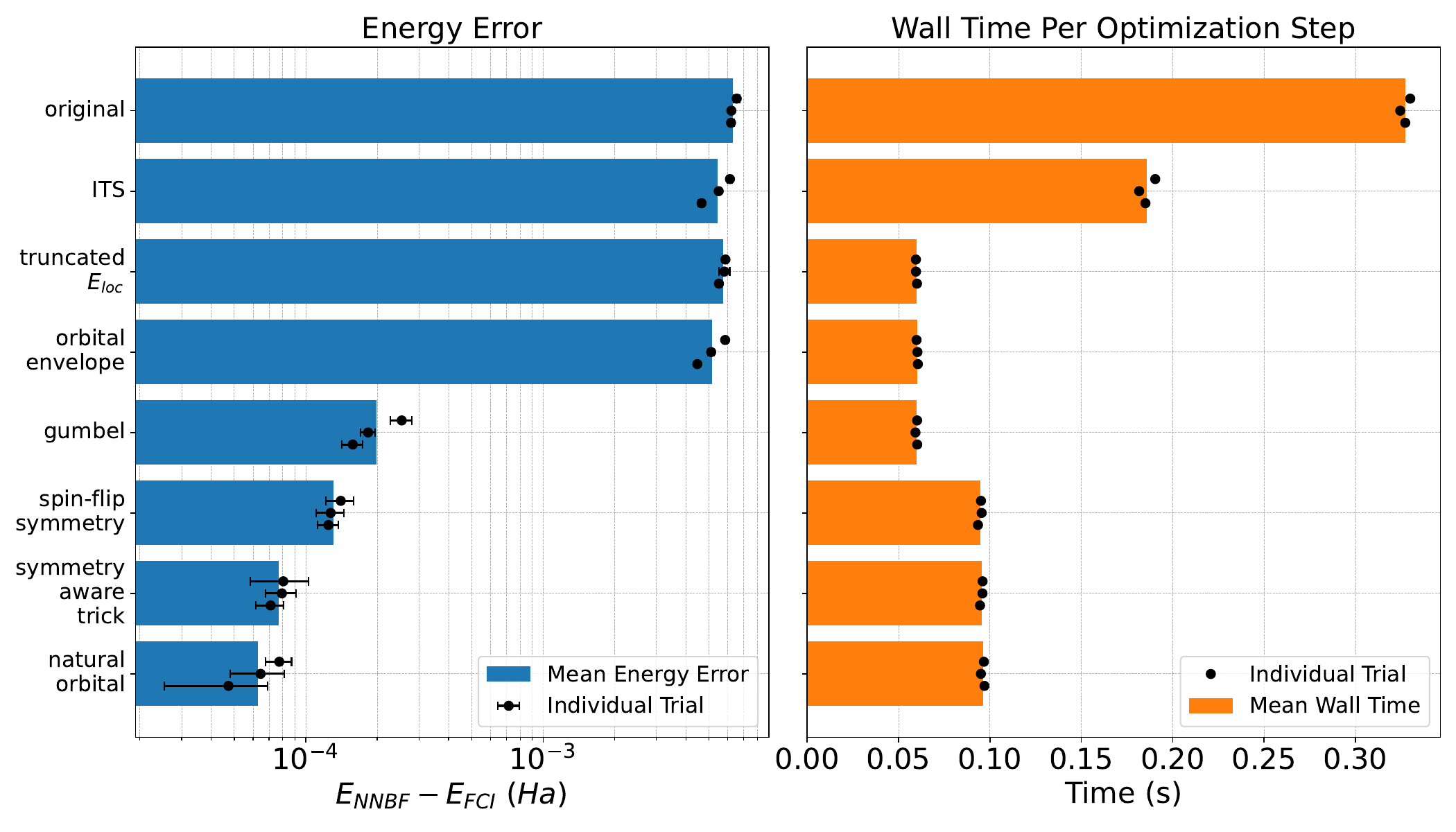}
    \caption{
    Evaluation of accuracy and speed improvements achieved through the algorithmic enhancements detailed in Section \ref{sec:gumbel}, \ref{sec:ITS}, \ref{sec:truncated_local_energy}, and \ref{sec:encode_physics}. Each bar represents the average energy error and per-optimization-step wall time over three independent runs, with the individual data points also shown. All calculations were performed on the Li$_2$O molecule with the STO-3G basis set, starting from canonical HF orbitals. The experiments used a fixed batch size of $|\mathcal{S}|=|\mathcal{V}|=1024$ and a network architecture of $(D,L,h)=(1,2,64)$. The algorithmic improvements are applied cumulatively.
    }
    \label{fig:ablation_study}
\end{figure*}

This subsection demonstrates how each algorithmic enhancement from Sections \ref{sec:ITS}, \ref{sec:truncated_local_energy}, \ref{sec:gumbel}, and \ref{sec:encode_physics} contributes to improvements in energy accuracy and computational efficiency. The analysis is performed via an ablation study on the Li$_2$O molecule (STO-3G basis, starting from HF orbitals). These enhancements are cumulatively added to our previous algorithm \cite{Liu2024} to highlight their individual and combined contributions, with key results illustrated in Figure \ref{fig:ablation_study}.

The most significant improvement in accuracy comes from introducing the Gumbel top-$k$ trick (Section \ref{sec:gumbel}). As shown in Figure \ref{fig:ablation_study}, this sampling method alone reduces the energy error by two orders of magnitude without imposing additional computational cost per step.

Efficiency is first enhanced by the Intermittent Target Selection (ITS) method (Section \ref{sec:ITS}). In contrast to our previous work \cite{Liu2024}, which used the entire connected space $\mathcal{C}$ as the target space, ITS constructs a much more compact subspace $\mathcal{U}$. For the settings studied, its size is reduced by a factor of $l \sim N_o-N_e$. This directly lowers the number of required amplitude evaluations, reducing the time per optimization step from 0.327 to 0.185 seconds for the system tested.

The training process is further accelerated by our truncated local energy strategy (Section \ref{sec:truncated_local_energy}). By replacing many computationally demanding neural network evaluations with efficient $O(\log{|\mathcal{U}|})$ lookups of precomputed amplitudes, this strategy significantly boosts computational efficiency. In our example, this modification reduced the per-optimization time from 0.185 to 0.060 seconds, without compromising energy accuracy.

Finally, incorporating physical knowledge into the model architecture (Section \ref{sec:encode_physics}) enhances the expressiveness of the NNBF ansatz, leading to another gain in energy accuracy. These features maintain the same asymptotic computational complexity. However, we note that enforcing spin-flip symmetry increases the wall time by a constant factor (approx. 1.58x in our tests), leaving the overall scaling unchanged. While the orbital envelope offers a marginal improvement for this specific system, we have found it is crucial for achieving high accuracy in larger molecules.

Collectively, these algorithmic enhancements enable the achievement of significantly improved energy accuracy and considerably reduced computational time for a given batch size and network architecture.

\subsubsection{The effect of Gumbel noise, inclusion probability, and renormalization}\label{sec:effect_gumbel_feature}
Given that Gumbel top-$k$ selection can significantly improve accuracy (as shown in Figure~\ref{fig:ablation_study}), it is instructive to investigate the contribution of its individual components. To isolate these contributions, we performed an ablation study on the Li$_2$O molecule (STO-3G basis, $|\mathcal{S}|=|\mathcal{V}|=1024$). We systematically enabled or disabled three key features of the sampling and weighting scheme: (1) the use of Gumbel noise to perturb selection probabilities, (2) the use of the inclusion probability $q_i(\kappa)$ [Eq.~\eqref{eq:inclusion_probability}] in the assignment of importance weights, and (3) the subsequent renormalization of these weights.

\begin{figure}[htbp!]
    \centering
    \includegraphics[width=\linewidth]{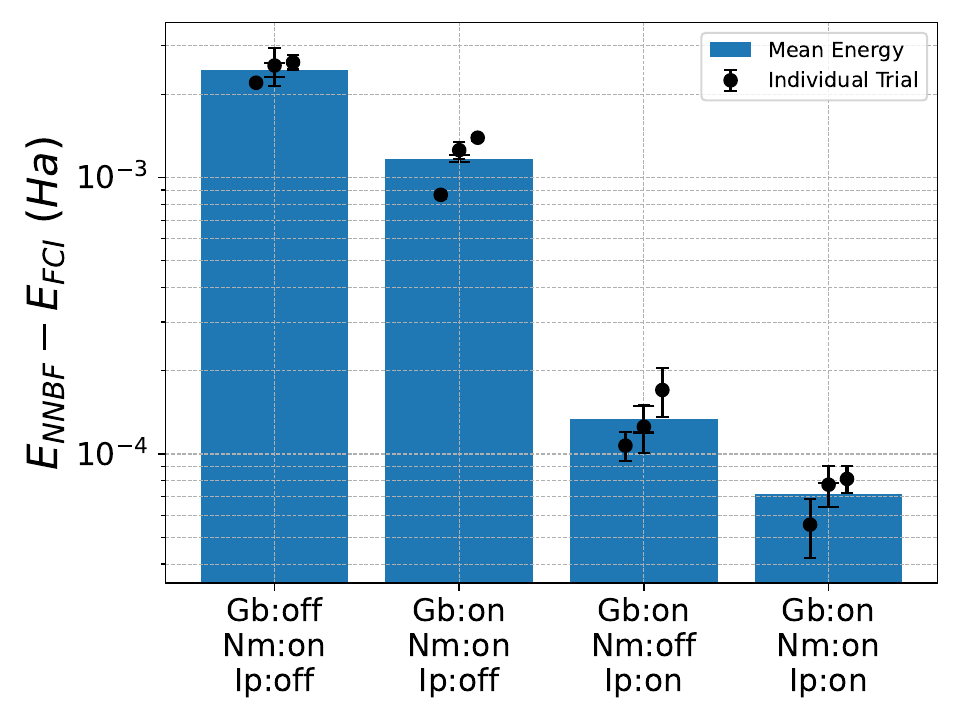}
    \caption{
    Ablation study showing the individual contribution of each component of the Gumbel top-$k$ sampling method (introduced in Section~\ref{sec:gumbel}) to the overall energy improvement detailed in Figure~\ref{fig:ablation_study}. All calculations were performed on the Li$_2$O molecule (STO-3G basis) with a $(D,L,h)=(1,2,64)$ network architecture and parameters $|\mathcal{S}|=|\mathcal{V}|=1024$. The labels on the x-axis indicate the components used in each experiment: \textbf{Gb} refers to the use of Gumbel noise during sample selection; \textbf{Ip} signifies that the importance weights incorporate the inclusion probability ($q_i(\kappa)$); and \textbf{Nm} denotes that the final weights are renormalized.
    }
    \label{fig:gumbel_investigation}
\end{figure}

The findings, presented in Figure~\ref{fig:gumbel_investigation}, demonstrate the distinct role of each component. Adding only Gumbel noise to the sampling process, without the corresponding unbiased reweighting, provides a marginal but noticeable advantage. This is expected, as the estimators for the energy and its gradient [Eqs.~\eqref{eq:exact_energy} and \eqref{eq:exact_gradient}] still depend entirely on the specific configurations in the sample $\mathcal{S}$. Conversely, incorporating the inclusion probability $q_i(\kappa)$ into the importance weights is crucial, reducing the energy error by an order of magnitude even without renormalization. Applying renormalization to these weights provides a final, noticeable improvement to the energy.

These observations align with the description in Section~\ref{sec:gumbel}. The inclusion probability provides an unbiased estimator over the entire target space $\mathcal{U}$, while the subsequent renormalization, in practice, improves the estimation by reducing variance, albeit at the potential cost of introducing a small bias.

\subsubsection{Local energy approximation strategy comparison}\label{sec:local_energy_strategy_comparison}

In Section~\ref{sec:truncated_local_energy}, we claimed that our truncated local energy strategy yields a more accurate approximation, and thus better training performance, than commonly used approaches at a comparable cost. Our method, denoted $E_l(\mathbf{x}_i|\mathcal{U})$, calculates the local energy for each sample $\mathbf{x}_i \in \mathcal{S}$ by summing over connected configurations within the entire target space $\mathcal{U}$. This contrasts with the common strategy, here denoted $E_l(\mathbf{x}_i|\mathcal{S})$, which restricts this sum to the much smaller sample set $\mathcal{S}$ itself \cite{Wu2023,Li2024,Malyshev2024}. The common approach typically also defines the importance weights as the amplitude-squared values renormalized within $\mathcal{S}$ (i.e., $w_i=|\psi(\mathbf{x}_i)|^2/\sum_{\mathbf{x}_j\in\mathcal{S}}{|\psi(\mathbf{x}_j)|^2}$). This combination of local energy truncation and weighting makes the training objective variational with respect to the sample set $\mathcal{S}$.

\begin{figure}[htbp!]
    \centering
    \includegraphics[width=\linewidth]{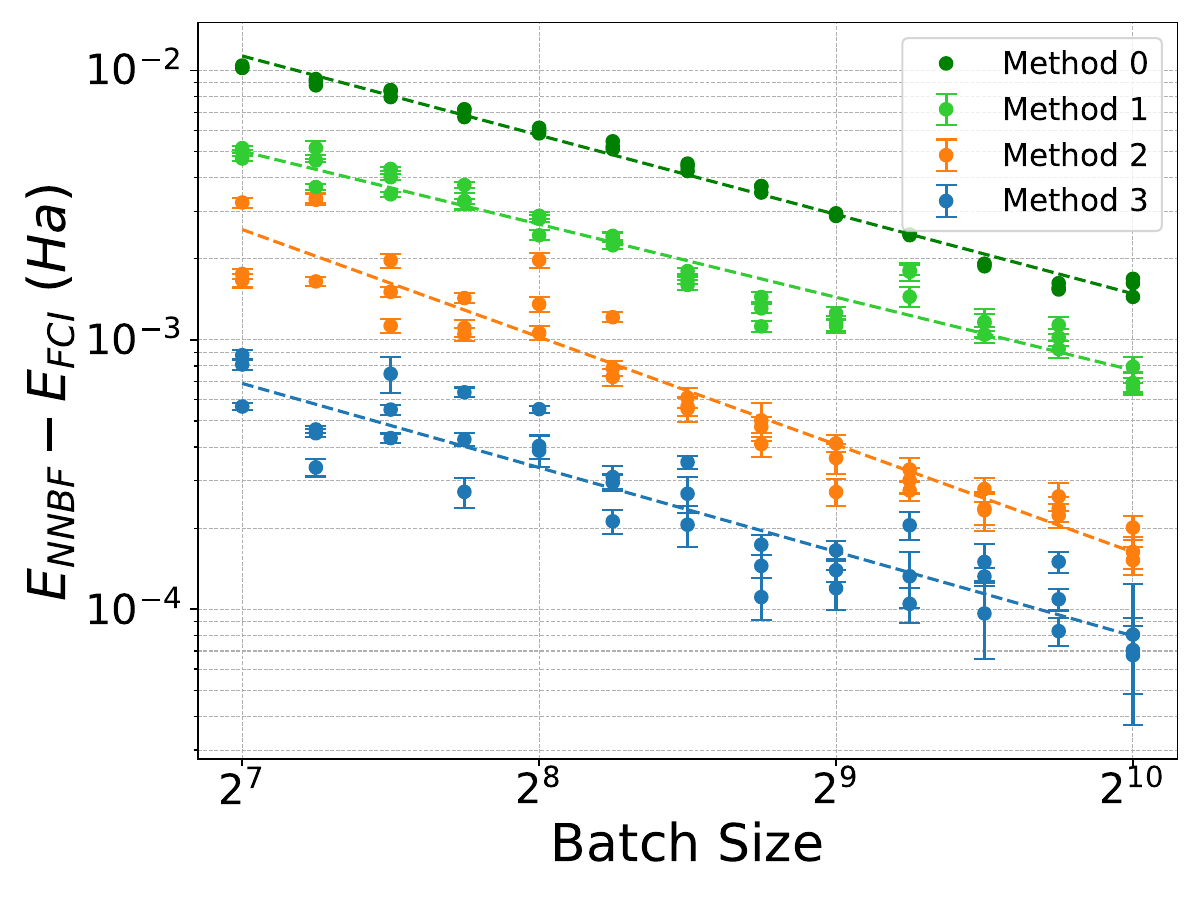}
    \caption{
    Cumulative performance comparison of different energy estimation strategies. All calculations were performed on the Li$_2$O molecule using the STO-3G basis set with canonical Hartree-Fock (HF) orbitals and a $(D,L,h)=(1,2,64)$ network architecture. The data points for each batch size correspond to three independent runs, and the accompanying lines are least-squares fits to this data. The four methods compared add features cumulatively. \textbf{Method 0} (baseline) uses a local energy sum over the sample set, $E_L(\mathbf{x}_i|\mathcal{S})$, with simple amplitude-squared weights, and reports the best training objective energy. \textbf{Method 1} uses the same training but reports a post-training MCMC inference energy. \textbf{Method 2} further improves the weights by incorporating the inclusion probability ($q_i(\kappa)$). \textbf{Method 3} (our full approach) combines these improved weights with our more accurate local energy calculation, $E_L(\mathbf{x}_i|\mathcal{U})$. The energy error of Method 3 decreases with the sample size $N=|\mathcal{S}|$ approximately following $N^{-1.038(69)}$ ($R^2=0.859$).
    }
    \label{fig:el_strategy}
\end{figure}

To substantiate our claim and isolate the sources of improvement, we performed a comparative study on the Li$_2$O molecule (STO-3G basis) across various batch sizes ($|\mathcal{S}|=|\mathcal{V}|$). We designed a cumulative comparison starting from a baseline that represents the common approach (Method 0): using $E_l(\mathbf{x}_i|\mathcal{S})$ with simple amplitude-squared weights and reporting the best objective function value from training. We then assess the impact of three modifications in sequence. First (Method 1), we leave the wave-function generated from Method 0 but  report the final energy using a global post-training MCMC inference which estimates the energy of the full wave-function. Second (Method 2), we additionally improve the training by incorporating the inclusion probability, $q_i(\kappa)$ [Eq. \eqref{eq:inclusion_probability}], into the importance weights. Finally (Method 3), our full approach combines these improvements with our more accurate local energy calculation, $E_l(\mathbf{x}_i|\mathcal{U})$. The results of this study are depicted in Figure~\ref{fig:el_strategy}.

The results in Figure~\ref{fig:el_strategy} reveal three key insights. First, comparing Method 1 to Method 0 shows that simply reporting the post-training MCMC inference energy provides a more accurate global estimate than using the best training objective value, reducing the error by an average factor of 2.12. This implies that the NNBF learns information about the wavefunction beyond the subspace it is directly trained on, reaffirming the representability and learnability of the ansatz. 

Second, comparing Method 2 to Method 1 demonstrates that introducing the inclusion probability to the importance weights significantly improves performance, reducing the energy error by a further factor of 3.23. This confirms our conclusion from Section \ref{sec:effect_gumbel_feature} that proper reweighting is crucial for improving the objective function estimation when using stochastic sampling without replacement. 

Third, comparing Method 3 (our full approach) to Method 2 shows that switching the local energy calculation from the sample set ($E_l(\mathbf{x}_i|\mathcal{S})$) to the target space ($E_l(\mathbf{x}_i|\mathcal{U})$) yields another non-trivial improvement, reducing the remaining error by an average factor of 2.94. This substantiates our central claim that leveraging the larger and more significant subspace $\mathcal{U}$ provides a more accurate local energy approximation, leading to superior overall performance. Importantly, we verified that the computational cost per optimization step is comparable for each method, confirming that these improvements in accuracy are not achieved at the expense of increased per-step complexity.

\subsection{NNBF Expressiveness and the Role of IPR}
Lastly, we examine the relationship between the inverse participation ratio (IPR) of the quantum state and the expressiveness of NNBF. Similar experiments have been conducted in Ref. \onlinecite{Malyshev2024} using autoregressive neural networks. In our study, we consider N$_2$ and CH$_4$ using the STO-3G basis set with HF orbitals, and use a vanilla NNBF ansatz—without employing spin-flip symmetry, spin-flip-symmetry-aware techniques, and orbital envelope features. To eliminate approximation errors in energy and its gradient, training is performed over the entire Hilbert space.

\begin{figure*}[htbp!]
  \centering

  \begin{subfigure}{0.49\linewidth}
    \includegraphics[width=\linewidth]{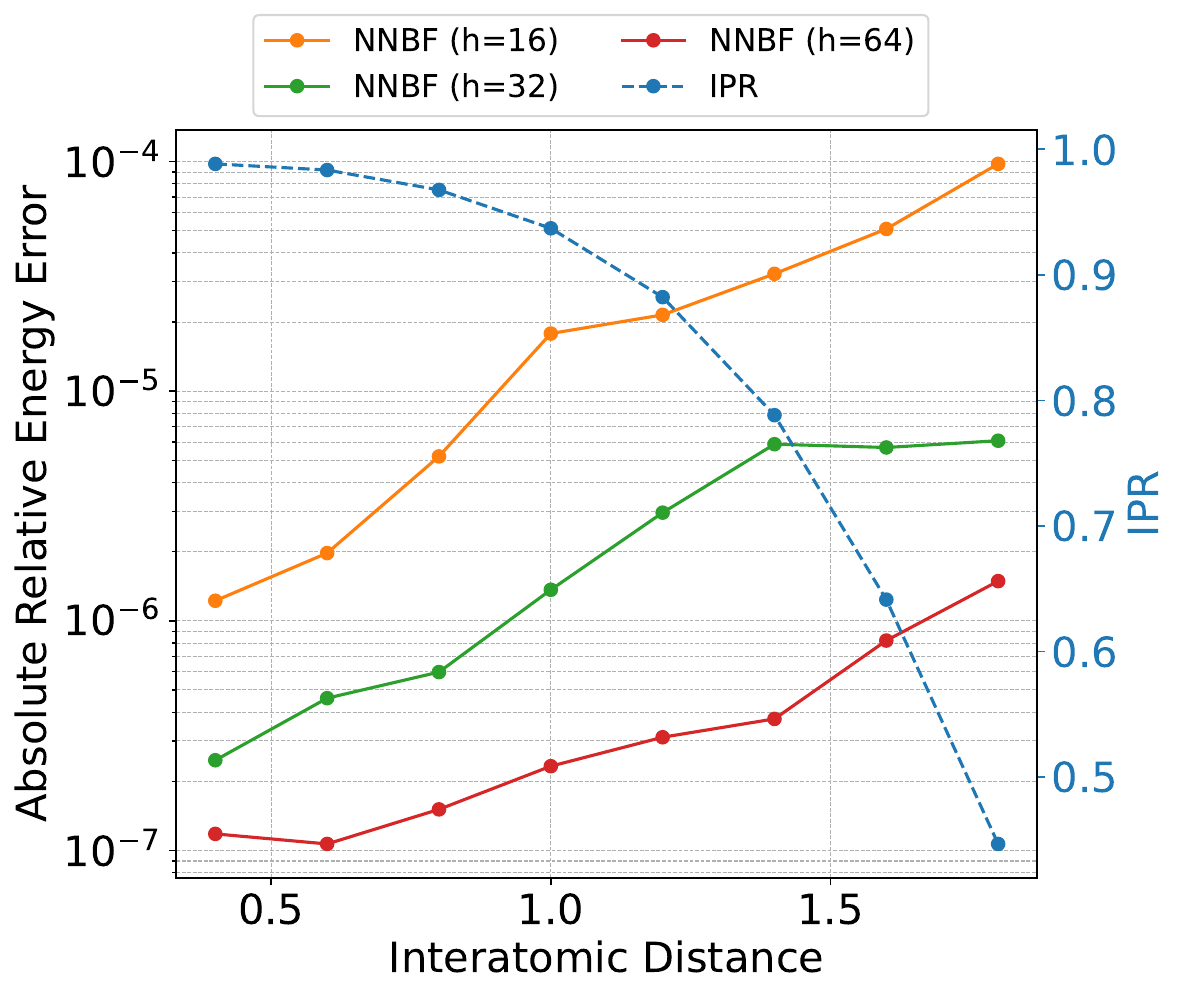}
    \caption{CH$_4$}
    \label{fig:IPR_CH4}
  \end{subfigure}
  \hfill
  \begin{subfigure}{0.49\linewidth}
    \includegraphics[width=\linewidth]{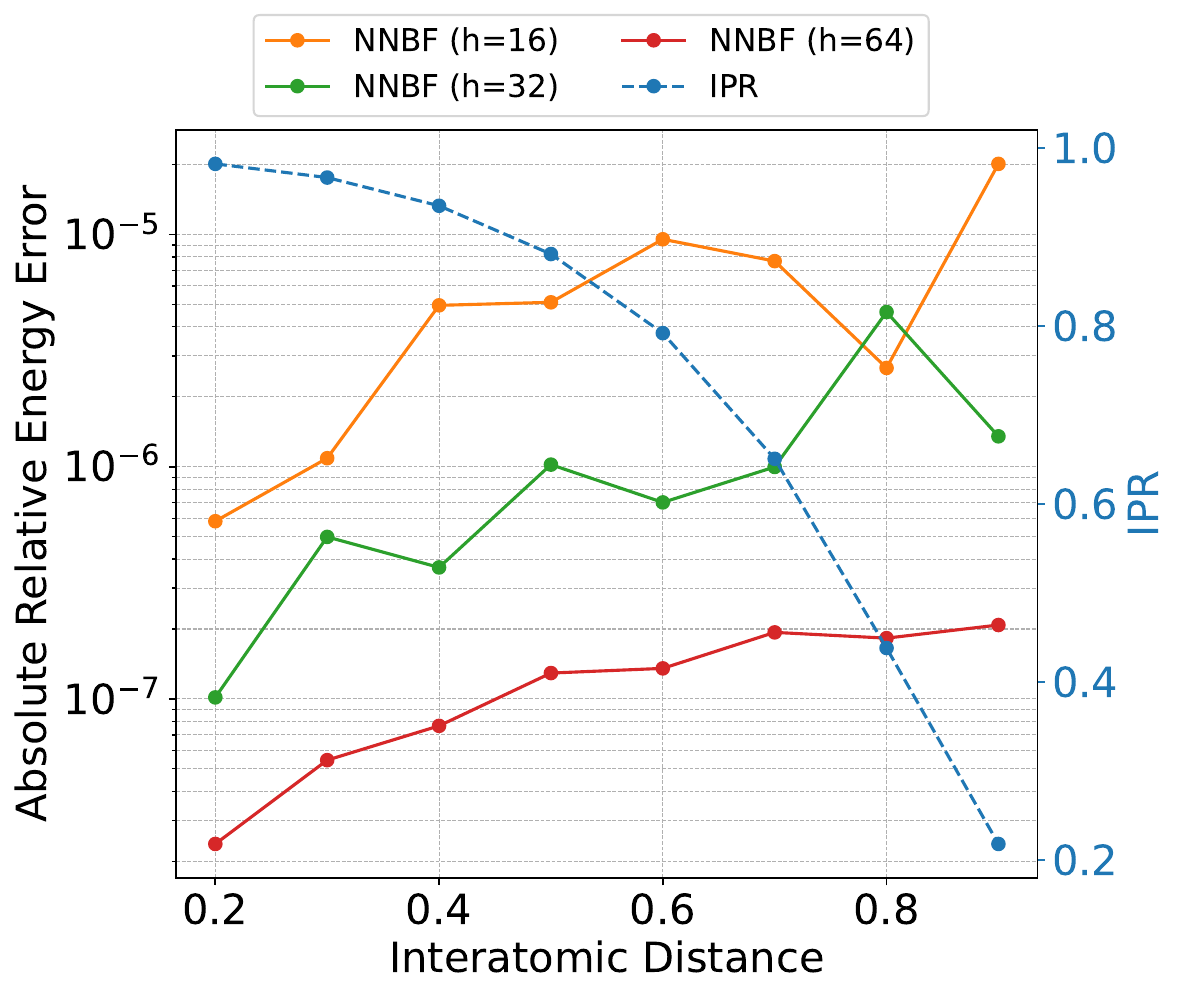}
    \caption{N$_2$}
    \label{fig:IPR_N2}
  \end{subfigure}
\caption{
Investigating the relationship between NNBF expressiveness and the inverse participation ratio (IPR) of the molecular ground state through the dissociation curves of (a) CH$_4$ and (b) N$_2$ molecules, using the STO-3G basis set with canonical HF orbitals. For each set of molecule geometry and network architecture, three trials are performed with different random seeds; each data point shows the smallest absolute relative energy error among them. The relative energy error is computed as the difference between the total NNBF energy and the total FCI energy, divided by the electronic FCI energy (excluding the fixed, non-variational nuclear repulsion contribution). The absolute relative energy error (left y-axis) and ground-state IPR (right y-axis) are plotted as functions of interatomic distance.  To eliminate approximation errors in energy and its gradient, optimization is performed over the entire Hilbert space ($\abs{\mathcal{S}}=\abs{\mathcal{H}}$). Double precision is employed for exact inference energy calculations, as higher precision becomes crucial when the NNBF state closely approximates the true ground state. Different hidden unit sizes are used to illustrate overparameterized ($h=64$) and underparameterized ($h=16,32$) scenarios.
}
\label{fig:IPR}
\end{figure*}

Figure \ref{fig:IPR} shows that the absolute relative energy error increases as the IPR decreases for both N$_2$ and CH$_4$ across both overparameterized ($h=64$) and underparameterized ($h=16$ and $h=32$) cases. While this observation is consistent with the intuition that highly peaked probability distributions are easier to optimize, other factors likely influence overall performance. For example, the complexity of the amplitude landscape, including the distribution of nodal regions and the interplay between electron correlation and orbital symmetries, might also contribute to these effects. Unraveling these factors represents an important direction for future research, and our findings offer valuable insights into the expressiveness of neural quantum states.

\section{Conclusions}
In this work, we have demonstrated that our proposed algorithmic enhancements—Intermittent Target Selection (ITS), truncated local energy evaluation, Gumbel top-$k$ selection, and physics-informed encoding—significantly improve the performance of the NNBF ansatz. Across a range of molecular systems, our method consistently achieves lower ground-state energies than existing NQS approaches and surpasses conventional CCSD and CCSD(T) calculations. The NNBF demonstrates its ability to capture both dynamic and static correlation by accurately matching the full dissociation curve of H$_2$O in both near-equilibrium and highly-stretched regimes. Furthermore, for the challenging N$_2$ molecule, our method achieves variational energies competitive with those from state-of-the-art SCI, FCIQMC, and DMRG calculations.

Our ablation studies quantify the impact of each new technique. We show that Gumbel top-$k$ sampling alone reduces the energy error by two orders of magnitude with no computational overhead, while the combination of ITS and our truncated local energy strategy significantly reduces the per-iteration wall time. We also present two additional focused comparative studies to provide deeper insight. The first dissects the Gumbel top-$k$ method, isolating the contributions from the stochastic noise, the inclusion probability weighting, and the final weight renormalization. The second study systematically compares our local energy strategy against the commonly used alternative, substantiating the benefits of our approach.

Lastly, our analysis of NNBF expressiveness shows that lower IPR values generally correspond to higher relative energy errors, indicating that more delocalized states are harder to approximate. Despite this trend, factors beyond IPR also influence the optimization difficulty, highlighting the complexity of the amplitude landscape.

Future work could focus on adopting more advanced optimizers, such as minSR methods \cite{Chen2024,Rende2024}, leveraging dynamic orbital rotations to produce more compact wavefunctions \cite{Tubman2020}, and integrating spin-flip symmetry directly into the NNBF’s internal neural network structure. We anticipate that the techniques developed in this study will enable more efficient and reliable NQS optimization, expanding the range of practical applications in quantum chemistry.


\begin{acknowledgments}
This work utilized the Illinois Campus Cluster, a computing resource operated by the Illinois Campus Cluster Program in collaboration with the National Center for Supercomputing Applications and supported by funds from the University of Illinois at Urbana-Champaign. We also acknowledge support from the NSF Quantum Leap Challenge Institute for Hybrid Quantum Architectures and Networks (NSF Award 2016136).
\end{acknowledgments}

\appendix
\section{Experimental Setup}\label{appx:experimental_setup}
\subsection{Training and Energy Inference Procedure}
The internal neural network used in this work is a multilayer perceptron (MLP) with $h$ hidden units, $L$ hidden layers, and $D$ backflow determinants, where residual connections are incorporated when $L>1$. We use the AdamW optimizer to minimize the energy expectation value of our NNBF ansatz and approximate the ground state of various molecules. The default hyperparameters are listed in Table \ref{tab:hyperparameters_and_notations}. Baseline energies for HF, CCSD, CCSD(T), and FCI calculations are obtained using the \textit{PySCF} software package \cite{pyscf}.

After training with the algorithmic enhancements described in Sections \ref{sec:ITS}, \ref{sec:truncated_local_energy}, \ref{sec:gumbel}, and \ref{sec:encode_physics}, some of which may be turned off in ablation studies, the energy expectation value and its statistical uncertainty are determined through post-training inference using a Markov Chain Monte Carlo (MCMC) procedure. We employ $N_w=1024$ concurrent walkers, each generating a Markov chain by sampling from the unnormalized probability distribution $\Bar{p}_\theta(\bm{x})=\psi_\theta(\bm{x})^2$ using the Metropolis–Hastings algorithm. Proposed moves consist of swapping an occupied spin-orbital with an unoccupied one of the same spin, and to reduce autocorrelation, the chains are downsampled at an interval of $K_1=10N_e$ iterations.

Before this sampling begins, the walkers are initialized to ensure robust exploration and avoid trapping in local minima, following the ensemble method outlined in ref.~\onlinecite{ForemanMackey2013}. Specifically, initial positions are drawn from a distribution defined by the eight most dominant configurations identified in the final stage of training. The walkers then undergo a burn-in period of $K_2=100K_1$ iterations to reach equilibrium.

Following the burn-in, we collect $T=1000$ configurations from each walker. The final energy is computed as the mean over the total $T \times N_w$ collected samples, and the reported uncertainty is the standard error of this mean, given by $\sqrt{\text{Var}(E)/(T \times N_w)}$. The energies reported in all experiments are obtained via this MCMC procedure, except when investigating the relationship between NNBF expressiveness and the IPR, where the energy is computed exactly over the full Hilbert space.

\begin{table*}[htbp!]
\centering
\begin{tabular}{lll}
\toprule
\textbf{Parameter / Symbol} & \textbf{Description} & \textbf{Default Value / Definition} \\
\midrule
\multicolumn{3}{l}{\textit{\textbf{A. Physical System \& Configuration Spaces}}} \\
\midrule
$N_e$ & Number of electrons & Problem-dependent \\
$N_o$ & Number of spin-orbitals & Problem-dependent \\
$\mathcal{H}$ & The physical Hilbert space & Problem-dependent \\
$\mathcal{V}$ & Core space & See Section \ref{sec:ITS} \\
$\mathcal{C}$ & Connected space & Generated from core space \\
$\mathcal{U}$ & Target space & See Section \ref{sec:ITS} \\
$\mathcal{S}$ & \makecell[l]{Sample set for energy estimations} & See Section \ref{sec:gumbel} \\
\midrule
\multicolumn{3}{l}{\textit{\textbf{B. Model Architecture (MLP)}}} \\
\midrule
$D$ & Number of backflow determinants & 1 \\
$L$ & Number of hidden layers & 2 \\
$h$ & Number of hidden units per layer & 256\\
\midrule
\multicolumn{3}{l}{\textit{\textbf{C. Optimizer \& Training Schedule}}} \\
\midrule
Optimizer & Algorithm for parameter updates & AdamW \\
$\beta_1, \beta_2$ & AdamW exponential decay rates & 0.9, 0.999 \\
$\epsilon$ & AdamW epsilon for numerical stability & $1 \times 10^{-8}$ \\
$\lambda$ & AdamW weight decay & $1 \times 10^{-4}$ \\
Learning Rate ($t$) & Initial rate with decay over iteration $t$ & $10^{-3} \times (1+10^{-5}t)^{-1}$ \\
Pretraining Iterations & Number of steps before main training & 500 \\
$N_w$ (pretraining) & Number of walkers during pretraining & 8192 \\
\midrule
\multicolumn{3}{l}{\textit{\textbf{D. Algorithmic Enhancements}}} \\
\midrule
$l$ & Speedup factor for ITS & $N_o - N_e$ \\
\midrule
\multicolumn{3}{l}{\textit{\textbf{E. Implementation Details}}} \\
\midrule
Framework & Core computational library & JAX \\
Precision & Floating-point precision & float32 \\
Energy Unit & Standard unit for energy values & Hartree \\
\midrule
\multicolumn{3}{l}{\textit{\textbf{F. Post-Training MCMC Inference}}} \\
\midrule
$M_{\text{init}}$ & Dominant configurations for initialization & 8 \\
$N_w$ (inference) & Number of MCMC walkers & 1024 \\
$K_2$ & Burn-in steps per walker & $100K_1$ \\
$K_1$ & Downsample interval (iterations) & $10N_e$ \\
$M$ & Samples collected per walker & 1000 \\
\bottomrule
\end{tabular}
\caption{Consolidated hyperparameters and notations used for all experiments, unless explicitly stated otherwise.}
\label{tab:hyperparameters_and_notations}
\end{table*}

\subsection{\texorpdfstring{Geometry and hyperparameters used for Section~\ref{sec:GSE_mol}}{Geometry and hyperparameters for ground-state calculations}}\label{appx:GSE_mol_geometry}

This section provides the specific molecular geometries and key training hyperparameters used for the ground-state energy calculations presented in Section~\ref{sec:GSE_mol}. The following table lists the Cartesian coordinates (in Angstroms) for each molecule. Alongside each geometry, we also specify the neural network architecture $(D,L,h)$ and the sizes of the sample and core spaces ($\abs{\mathcal{S}}$ and $\abs{\mathcal{V}}$) used for that particular calculation. These values, in conjunction with the general methodology and default parameters outlined in Appendix~\ref{appx:experimental_setup}, define the complete setup for each experiment.

\begin{table}[htbp!]
\centering
\begin{tabular}{cccc}
\toprule
\textbf{Atom} & \textbf{x} & \textbf{y} & \textbf{z} \\
\midrule
\multicolumn{4}{c}{\textbf{N$_2$} --- $(D,L,h)=(1,2,256)$, $|\mathcal{S}|=|\mathcal{V}|=1024$} \\
\midrule
N & -0.556 &  0.0 &  0.0 \\
N &  0.556 &  0.0 &  0.0 \\
\midrule
\multicolumn{4}{c}{\textbf{CH$_4$} --- $(D,L,h)=(1,2,256)$, $|\mathcal{S}|=|\mathcal{V}|=1024$} \\
\midrule
C &  0.0 &  0.0 &  0.0 \\
H &  0.5541 &  0.7996 &  0.4965 \\
H &  0.6833 & -0.8134 & -0.2536 \\
H & -0.7782 & -0.3735 &  0.6692 \\
H & -0.4593 &  0.3874 & -0.9121 \\
\midrule
\multicolumn{4}{c}{\textbf{LiF} --- $(D,L,h)=(1,2,256)$, $|\mathcal{S}|=|\mathcal{V}|=1024$} \\
\midrule
F &  2.0 &  0.0 &  0.0 \\
Li &  3.0 &  0.0 &  0.0 \\
\midrule
\multicolumn{4}{c}{\textbf{CH$_2$O} --- $(D,L,h)=(1,2,256)$, $|\mathcal{S}|=|\mathcal{V}|=1024$} \\
\midrule
O &  0.6123 &  0.0 &  0.0 \\
C & -0.6123 &  0.0 &  0.0 \\
H & -1.2 &  0.2426 & -0.8998 \\
H & -1.2 & -0.2424 &  0.8998 \\
\midrule
\multicolumn{4}{c}{\textbf{LiCl} --- $(D,L,h)=(1,2,256)$, $|\mathcal{S}|=|\mathcal{V}|=1024$} \\
\midrule
Cl &  2.0 &  0.0 &  0.0 \\
Li &  3.0 &  0.0 &  0.0 \\
\midrule
\multicolumn{4}{c}{\textbf{CH$_4$O} --- $(D,L,h)=(1,2,256)$, $|\mathcal{S}|=|\mathcal{V}|=1024$} \\
\midrule
O &  0.7079 &  0.0 &  0.0 \\
C & -0.7079 &  0.0 &  0.0 \\
H & -1.0732 & -0.769 &  0.6852 \\
H & -1.0731 & -0.1947 & -1.0113 \\
H & -1.0632 &  0.9786 &  0.3312 \\
H &  0.9936 & -0.8804 & -0.298 \\
\midrule
\multicolumn{4}{c}{\textbf{Li$_2$O} --- $(D,L,h)=(1,2,256)$, $|\mathcal{S}|=|\mathcal{V}|=1024$} \\
\midrule
O &  2.866 & -0.25 &  0.0 \\
Li &  3.732 &  0.25 &  0.0 \\
Li &  2.0 &  0.25 &  0.0 \\
\midrule
\multicolumn{4}{c}{\textbf{C$_2$H$_4$O} --- $(D,L,h)=(1,2,512)$, $|\mathcal{S}|=|\mathcal{V}|=4096$} \\
\midrule
O & -0.0007 &  0.8141 &  0.0 \\
C &  0.7509 & -0.4065 &  0.0 \\
C & -0.7502 & -0.4076 &  0.0 \\
H &  1.2625 & -0.6786 &  0.9136 \\
H &  1.2625 & -0.6787 & -0.9136 \\
H & -1.2614 & -0.6806 & -0.9136 \\
H & -1.2614 & -0.6805 &  0.9136 \\
\midrule
\multicolumn{4}{c}{\textbf{C$_2$H$_4$O$_2$} --- $(D,L,h)=(1,2,512)$, $|\mathcal{S}|=32768, |\mathcal{V}|=4096$} \\
\midrule
O & -0.3035 &  1.289 & -0.0002 \\
O & -0.98 & -0.8878 & -0.0002 \\
C &  1.3743 & -0.3516 & -0.0002 \\
C & -0.0907 & -0.0496 &  0.0006 \\
H &  1.8368 &  0.057 & -0.9021 \\
H &  1.84 &  0.0676 &  0.8952 \\
H &  1.5207 & -1.4356 &  0.0064 \\
H & -1.2598 &  1.5081 & -0.0008 \\
\bottomrule
\end{tabular}
\caption{Geometries (in Angstroms) and corresponding training hyperparameters for all molecules studied in Section~\ref{sec:GSE_mol}.}
\label{tab:geometries_consolidated}
\end{table}

\section{Learned Orbital Envelope Parameters}\label{appx:learned_orbital_envelope}
\begin{figure}[htbp!]
\centering
\includegraphics[width=\linewidth]{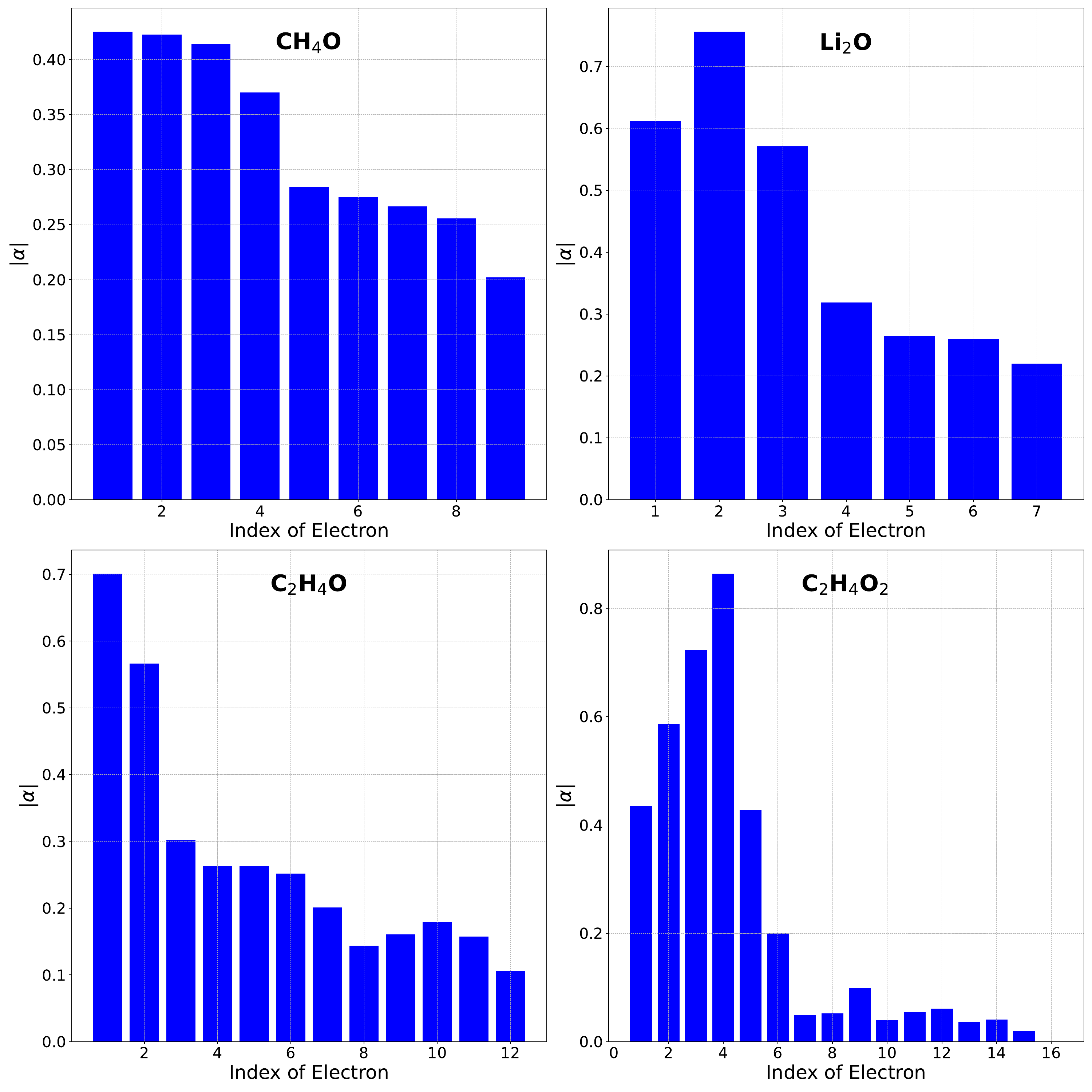}
\caption{Distribution of the orbital envelope parameters $\alpha_i$ after training for CH$_4$O, Li$_2$O, C$_2$H$_4$O, and C$_2$H$_4$O$_2$ in the STO-3G basis with CCSD natural orbitals. The y-axis shows the absolute value of $\alpha_i$, as only its magnitude affects the envelope per Eq. \eqref{eq:orbital_envelope}.}
\label{fig:distribution_orbital_envelope}
\end{figure}

This section illustrates what the orbital envelope [Eq. \eqref{eq:orbital_envelope}] learns during training by visualizing the distribution of its trainable parameters for the last four molecules listed in Table \ref{tab:molecule_energy}. All $\alpha$ values are initially set to a small value (0.01) to prevent biasing the ansatz. As shown in Figure \ref{fig:distribution_orbital_envelope}, after training, the $\alpha$ values associated with inner electrons become noticeably larger than those for outer electrons, reflecting the intuitive physical expectation that inner electrons remain closer to the nucleus in lower-energy orbitals.

\clearpage 
\bibliography{reference}

\end{document}